\documentclass[aps,prd,nofootinbib,twocolumn,superscriptaddress,preprintnumbers,balancelastpage,longbibliography]{revtex4-1}

\usepackage[utf8]{inputenc}
\usepackage{amsmath,amssymb,mathtools,bm}
\usepackage{graphicx, color, hepunits}
\usepackage[dvipsnames]{xcolor}
\usepackage{float}
\usepackage{multirow}
 \usepackage{hyperref} 
\hypersetup{
    colorlinks=true,      
    linkcolor=blue,        
    citecolor=blue,        
    filecolor=magenta,     
    urlcolor=blue          
}
\usepackage[utf8]{inputenc}
\usepackage[english]{babel}
\usepackage{lipsum}
\usepackage{mathrsfs}

\usepackage{tensor}

\newcommand*\diff{\mathop{}\!\mathrm{d}}

\def\beq{\begin{equation}}
\def\eeq{\end{equation}}

\def\Beq{\begin{equation}\begin{aligned}}
\def\Eeq{\end{aligned}\end{equation}}

\def\bea{\begin{eqnarray}}
\def\eea{\end{eqnarray}}

\def\beq{\begin{equation}}
\def\eeq{\end{equation}}
\def\bea{\begin{eqnarray}}
\def\eea{\end{eqnarray}}

\def\bp{\boldsymbol{p}}

\def\bk{\boldsymbol{k}}

\def\bx{\boldsymbol{x}}
\def\bq{\boldsymbol{q}}

\begin{document}

\preprint{DESY-23-065}

\title{A New Window into Gravitationally Produced Scalar Dark Matter}

\author{Marcos A. G. Garcia}
\affiliation{Departamento de F\'isica Te\'orica, Instituto de F\'isica, Universidad Nacional Aut\'onoma de M\'exico, Ciudad de M\'exico C.P. 04510, Mexico}

\author{Mathias Pierre}
\affiliation{Deutsches Elektronen-Synchrotron DESY, Notkestr. 85, 22607 Hamburg, Germany} 

\author{Sarunas Verner}
\affiliation{Institute for Fundamental Theory, Physics Department, University of Florida, Gainesville, FL 32611, USA}
\vspace{0.5cm}

\date{\today}

\begin{abstract}
Conventional scenarios of purely gravitationally produced dark matter with masses below the Hubble parameter at the end of inflation are in tension with Cosmic Microwave Background (CMB) constraints on the isocurvature power spectrum. We explore a more general scenario with a nonminimal coupling between the scalar dark matter field and gravity, which allows for significantly lighter scalar dark matter masses compared to minimal coupling predictions. By imposing relic abundance, isocurvature, Lyman-$\alpha$, and Big Bang Nucleosynthesis (BBN) constraints, we show the viable parameter space for these models. Our findings demonstrate that the presence of a nonminimal coupling expands the parameter space, yielding a dark matter mass lower bound of $2 \times 10^{-4} \, \rm{eV}$.
\end{abstract}
\maketitle

\noindent
{\bf Introduction.---} 
The nature and origin of dark matter (DM) remain one of the greatest unsolved mysteries in fundamental physics. Galaxy rotation curve observations, cosmic structure analyses, and gravitational lensing studies~\cite{ParticleDataGroup:2022pth, Rubin:1970zza} contribute to our understanding of dark matter, but reveal minimal information about its inherent characteristics. Furthermore, the lack of detection from indirect DM and terrestrial experiments, coupled with the stringent limitations imposed by direct DM detection searches such as XENON1T~\cite{XENON:2018voc}, LUX~\cite{LUX:2016ggv}, PandaX~\cite{PandaX-4T:2021bab}, and LZ~\cite{LZ:2022ufs}, challenges the conventional weakly interacting massive particle (WIMP) paradigm without providing new insights into the composition of the universe's invisible component. This discrepancy necessitates exploring alternative dark matter models~\cite{Arcadi:2017kky,Escudero:2016gzx,Mambrini}.\\
\indent One of the most well-motivated and notably minimalistic models involves the gravitational production of hidden sector particles during the transition from the inflationary quasi-de Sitter phase to a matter-dominated (MD) or radiation-dominated (RD) universe~\cite{Parker:1968mv, Parker:1969au, Ford:1986sy, Ema3, Chung:1998rq}. During the reheating phase, when the inflaton coherently oscillates about a minimum, the space-time curvature
also oscillates, leading to additional particle production~\cite{Garcia:2022vwm, Ema1, Ema2}. However, for nearly 20 years, it has been known that the CMB bounds on the amplitude of the isocurvature power spectrum imply that the purely gravitationally produced dark matter must be superheavy, i.e.~close to the Hubble scale at the end of inflation~\cite{Chung:2004nh, Chung:2011xd, Chung:2015pga, Herring:2019hbe, Padilla:2019fju, Ling:2021zlj, Redi:2022zkt}. In the present work, we demonstrate that when the spectator dark matter field couples nonminimally to gravity~\cite{Cosme:2017cxk,Cosme:2018nly, Alonso-Alvarez:2018tus, Fairbairn:2018bsw, Alonso-Alvarez:2019ixv, Kolb:2022eyn}, such models avoid the isocurvature constraints~\cite{Chung:2004nh, Cosme:2017cxk}, opening up the parameter space even for light dark matter with masses significantly below the Hubble scale at the end of inflation. These models exhibit numerous notable properties, and for the first time, a detailed consideration of the purely gravitational production and how the dark matter phase space distribution (PSD) changes in the presence of nonminimal coupling is presented. By imposing relic abundance, isocurvature, Big Bang Nucleosynthesis, and Lyman-$\alpha$ limits, we constrain the values of the nonminimal coupling and demonstrate that it opens up a significant portion of the allowed spectator dark matter parameter space.

\noindent
{\bf Framework.---} 
In this Letter, we use natural units $k_B = \hbar = c = 1$ and the signature $(+, -, -, -)$ for the spacetime metric $g_{\mu \nu}$.\footnote{We use the $(-,+,+)$ convention for the metric, Riemann tensor and Einstein equation according to the Misner-Thorne-Wheeler classification~\cite{Misner:1973prb}.} We consider the homogeneous and isotropic Friedmann-Robertson-Walker (FRW) metric $\diff s^2 = a(\eta)^2(\diff \eta^2 - \delta_{ij}\diff x^i \diff x^j)$, where $a(\eta)$ represents the scale factor and $\diff \eta = \diff t/a$ is the conformal time. The general action of our theory is given by
\begin{align}  
    \label{eq:action}
    \mathcal{S} \; = \; &\int \diff^4 x \sqrt{-g} \left[-\frac{1}{2}(M_P^2 - \xi \chi^2)R \right.\nonumber\\
    &+ \frac{1}{2} (\partial_{\mu} \phi)^2 - V(\phi) 
    \left. + \frac{1}{2} (\partial_{\mu} \chi)^2 - \frac{1}{2}m_{\chi}^2 \chi^2 \right]  \, .
\end{align}
Here $g = \det g_{\mu \nu}$ represents the determinant of the metric, $M_P = 1/\sqrt{8 \pi G_N} \simeq 2.435 \times 10^{18} \, \rm{GeV}$ denotes the reduced Planck mass, $R$ is the Ricci scalar, $\xi$ is the nonminimal coupling of the dark matter field to gravity, with $\xi = 0$ and $\xi = 1/6$ corresponding to minimal and conformal couplings, respectively. $\phi$ is the inflaton field, where $V(\phi)$ is the corresponding potential, and $\chi$ is the spectator scalar dark matter field whose bare mass is denoted by $m_\chi$. The $Z_2$ symmetry of the dark matter potential ensures its stability.

Introducing the canonically normalized field 
$X \equiv a \chi$, and varying the action~(\ref{eq:action}) with respect to $X$, we obtain the equation of motion 
\begin{equation}
    \label{eq:eom}
    (\partial_{\eta}^2 - \nabla^2 + a^2 m_{\rm{eff}}^2)X = 0,~~~~~m_{\rm{eff}}^2 = m_{\chi}^2 + \frac{1}{6}(1-6\xi)R \, .
\end{equation}
During inflation, one can approximate the Ricci scalar as equal to its de Sitter value, $R \simeq - 12H^2$, with $H$ the Hubble parameter, and the effective mass with minimal coupling ($\xi = 0$) becomes $m_{\text{eff}}^2 = m_{\chi}^2 - 2H^2$, whereas with conformal coupling ($\xi = 1/6$), it becomes $m_{\rm{eff}}^2 = m_{\chi}^2$. This implies that light scalars $m_\chi \ll H$ minimally coupled to gravity would experience a tachyonic phase
during inflation with $m_{\text{eff}}^2 < 0$. Problematically, 
during inflation, the tachyonic instability of light DM modes can efficiently generate isocurvature perturbations at the second order and lead to an unsuppressed isocurvature spectrum for long-wavelength (IR) modes~\cite{Chung:2004nh, Chung:2011xd, Ling:2021zlj, Garcia:2023awt}, in disagreement with the current isocurvature power spectrum constraints from \textit{Planck} observations~\cite{Planck:2018jri}. Nevertheless, when the conformal coupling is sufficiently large, one can approximate the effective mass as $m_{\text{eff}}^2 \simeq 12 \xi H^2$, which implies that the dark matter effective mass is very large during inflation but becomes much smaller at the end of inflation when the Hubble parameter drops to lower values, successfully avoiding the current isocurvature bound.

Since the FRW metric is spatially 
homogeneous, one can perform a Fourier expansion of the dark matter field $X$:
\begin{equation}
\label{eq:modedef}
X(\eta,\bx)  = \int \frac{\diff^3\bk}{(2\pi)^{3/2}}\,e^{-i\bk\cdot\bx} \left[ X_k(\eta)\hat{a}_{\bk} + X_k^*(\eta)\hat{a}^{\dagger}_{-\bk} \right]\,, 
\end{equation}
where $\bk$ is the comoving momentum, with  $|\bk| = k$, and $\hat{a}_{\bk}$ and $\hat{a}^{\dagger}_{\bk}$ are the annihilation and creation operators, respectively, which obey the canonical commutation relations $[\hat{a}_{\bk},\hat{a}^{\dagger}_{\bk'} ] = \delta(  \bk-\bk')$ and $[\hat{a}_{\bk},\hat{a}_{\bk'} ] = [\hat{a}^{\dagger}_{\bk},\hat{a}^{\dagger}_{\bk'} ] = 0$. The canonical commutation relations between the field, $X_k$, and its momentum conjugate, $X_k'$, are satisfied if the Wronskian condition $X_k X^{*\prime}_k - X_k^*X_k' \;=\; i$
is imposed. If we substitute the Fourier decomposed field~(\ref{eq:modedef}) into the equation of motion~(\ref{eq:eom}), we find that the equation of motion together with the angular frequency are given by 
\begin{equation}
    \label{eq:modes}
    X_k'' + \omega_k^2 X_k \; = \; 0 \, , \quad {\rm{with}} \quad \omega_k^2 \; = \; k^2 + a^2 m_{\rm{eff}}^2 \, .
\end{equation}
\noindent{\bf Gravitational Production of Dark Matter.---} To compute the gravitational production of dark matter during inflation and reheating stages, one must specify the initial conditions and solve the mode equations~(\ref{eq:modes}). In the early-time asymptotic limit, $\eta \rightarrow -\infty$, it leads to a flat Minkowski space, with $R \rightarrow 0$ and $a \rightarrow 0$. Thus, in this limit, the modes that satisfy $a H \ll k$ are deep inside the horizon, and the mode frequency can be approximated as $\omega_k = k$. This motivates the choice of the Bunch-Davies vacuum initial condition, given by $\lim \limits_{\eta \rightarrow - \infty} X_k(\eta) = \frac{1}{\sqrt{2k}} e^{-i k \eta}$. The comoving number density of gravitationally produced scalar dark matter, $n_{\chi}$, can be computed using the following expression~\cite{Kofman:1997yn, Garcia:2021iag, Ling:2021zlj}:
\begin{equation}
    \label{eq:comovingnd}
    n_{\chi} \left(\frac{a}{a_{\rm{end}}} \right)^3 \; = \; \int_{k_0}^{\infty} \frac{\diff k}{k} \mathcal{N}_k~~{\rm{with}}~~\mathcal{N}_k \; = \; \frac{k^3}{2\pi^2} f_{\chi}(k, t) \, ,
\end{equation}
where $a_{\rm{end}}$ is the scale factor at the end of inflation, and $\mathcal{N}_k$ is the comoving number density spectrum expressed as a function of the DM phase space distribution, $f_{\chi}(k,t) =  \frac{1}{2\omega_k}\left| \omega_k X_k - i X'_k \right|^2$.\footnote{The DM phase space distribution can also be expressed in terms of the Bogoliubov coefficients $f_{\chi}(k, t) = |\beta_k|^2 = \mathcal{N}_k$.} Here, we have introduced an IR cutoff, $k_0 = a_0 H_0$, where $a_0 = a(\eta_0)$ is the present-day scale factor and $H_0 = H(\eta_0)$ represents the present-day Hubble parameter. We note that $k_0$ represents the present comoving scale, assuming that inflation started when this mode was inside the horizon. Modes with lower wavenumbers are outside of our cosmological horizon and, as a result, contribute to the homogeneous background~\cite{Starobinsky:1994bd}. It can be shown both analytically and numerically that the phase space distribution in the long-wavelength (IR) regime scales as $f_{\chi} \propto k^{-2 \nu}$ ($\mathcal{N}_k \propto k^{3 - 2\nu}$), where $\nu = \sqrt{9/4 - 12\xi - m_{\chi}^2/H^2}$ for real $\nu$. If we assume that the dark matter scalar is light, with $m_{\rm{\chi}} \ll H$, we find that $f_{\chi} \propto k^{-3}$ ($\mathcal{N}_k = \rm{const}.$, flat spectrum) for minimal coupling and $f_{\chi} \propto k^{-1}$ ($\mathcal{N}_k \propto k^2$) for conformal coupling~\cite{Herring:2019hbe, Ling:2021zlj, Kaneta:2022gug, Garcia:2022vwm}. The comoving number density~(\ref{eq:comovingnd}) contains an IR divergence for $\xi = 0$, 
which is regulated by $k_0$. However, the integral is convergent for $\xi > 0$ and becomes IR insensitive approximately when $\xi \gtrsim 1/5$, which is close to the conformal coupling value $\xi = 1/6$.

\begin{figure}[!t]
\includegraphics[width = 1\columnwidth]{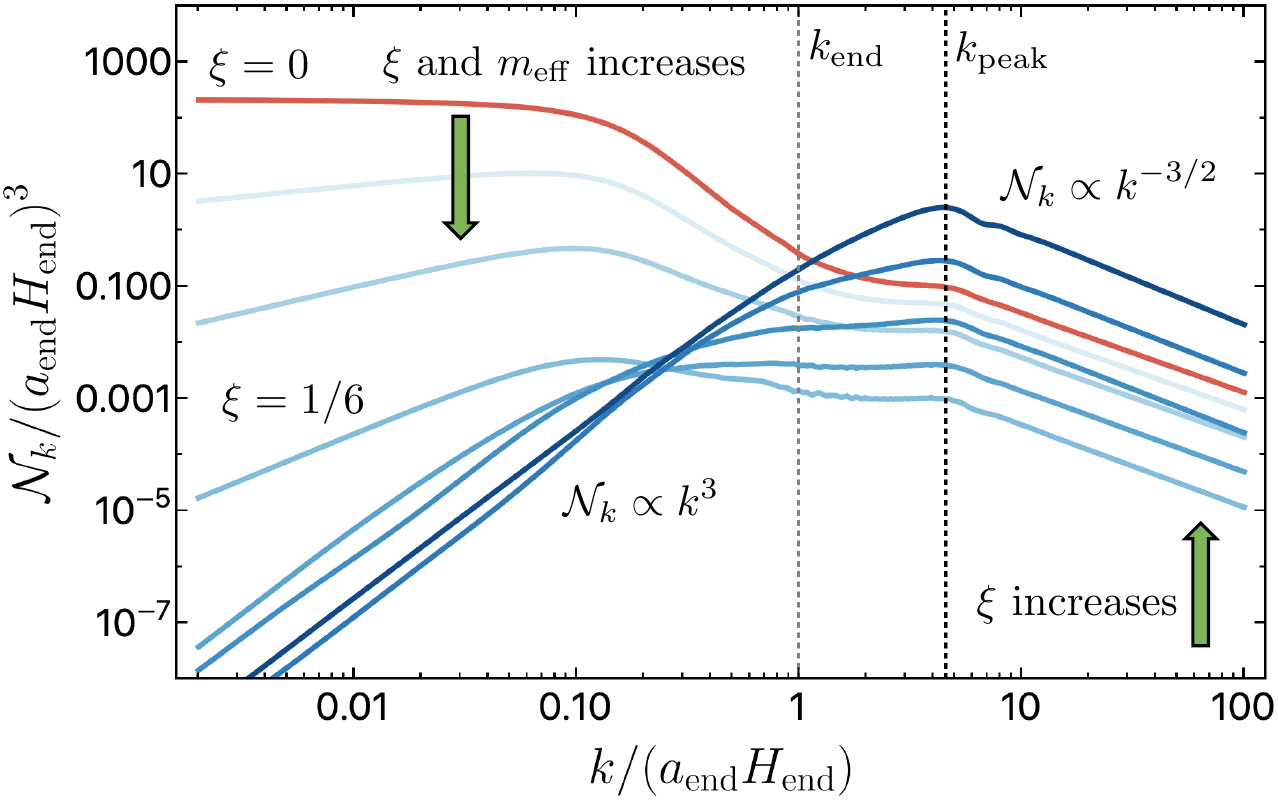}
\caption{A qualitative diagram illustrating the dependence of the comoving number density spectrum $\mathcal{N}_k$ on nonminimal coupling $\xi$ as a function of rescaled horizon modes $k/(a_{\rm{end}} H_{\rm{end}})$.}
\label{fig:genpsd}
\end{figure} 

We illustrate the characteristic dependence of the comoving number density spectrum on $\xi$ in Fig.~\ref{fig:genpsd}. We note that $\mathcal{N}_k$ typically peaks at $k_{\rm{peak}} > k_{\rm{end}}$, where $k_{\rm{end}} = a_{\rm{end}} H_{\rm{end}}$ is the mode that reenters the horizon at the end of inflation. 

The mode $k_{\rm{peak}}$, and the short-wavelength (UV) tail of the spectrum, correspond to modes that remain inside the horizon during inflation. For $k>k_{\rm preak}$, the spectrum scales as $f_{\chi} \propto k^{-9/2}$ ($\mathcal{N}_k \propto k^{-3/2}$), independently of the  value of $\xi$. The amplitude of $\mathcal{N}_k$ at $k_{\rm{peak}}$ increases with increasing value of the coupling $\xi$. 

For a large coupling $\xi$, the effective mass (\ref{eq:eom}) increases, along with the parameter $\nu$, and the spectrum becomes suppressed in the IR. The same effect occurs in the case of direct DM-inflaton coupling~\cite{Garcia:2022vwm, Garcia:2023awt}. The dominant contribution to gravitational particle production occurs when the standard adiabaticity condition is violated, $\dot{\omega}_k (m_{\rm{eff}}) \gtrsim \omega_k^2 (m_{\rm{eff}})$. When the coupling $\xi$ is very large, $m_{\rm{eff}} \simeq 12 \xi H^2$, and the transition from quasi-dS to a matter- or radiation-dominated universe causes the effective mass to change rapidly, leading to substantial particle production. Furthermore, when $\xi$ becomes very large (and $\nu$ is imaginary), we find that $f_{\chi} = \rm{const}.$ ($\mathcal{N}_{k} \propto k^{3}$). 

To determine the present DM relic abundance, one must compute the comoving number density~(\ref{eq:comovingnd}) beyond the end of the reheating epoch. We define reheating to occur at time $t_{\rm{reh}}$, when the inflaton energy density equals the energy density of the produced radiation, $\rho_{\phi}(t_{\rm{reh}}) = \rho_r(t_{\rm{reh}})$. The reheating temperature is then computed through the corresponding expression~\cite{Garcia:2020wiy}
\begin{equation}
    \label{eq:reheating}
    \rho_r(t_{\rm{reh}}) \; = \; \frac{\pi^2 g_{\rm{reh}} T_{\rm{reh}}^4}{30} \; \equiv \; \frac{12}{25} \left(\Gamma_{\phi} M_P \right)^2 \, ,
\end{equation}
where $g_{\rm{reh}}$ represents the effective number of relativistic degrees of freedom at reheating, $T_{\rm{reh}}$ is the reheating temperature, and $\Gamma_{\phi} \equiv \; y^2/(8 \pi) m_{\phi} $ is the inflaton perturbative decay rate, with $y$ corresponding to an effective Yukawa-like coupling and $m_{\phi}$ is the inflaton mass.

Assuming that reheating occurs significantly after the end of inflation, with $a_{\rm{reh}} \gg a_{\rm{end}}$, and that the total entropy $S=sa^{3}$ is conserved after the end of reheating, and using Eq.~(\ref{eq:comovingnd}), we find that the DM relic density can be expressed as~\cite{Garcia:2022vwm}
\begin{equation}
\Omega_{\rm DM, \, grav} \;\simeq\; \frac{m_{\chi}n_{\chi}}{\rho_c}=\; \frac{1}{6\pi q_0^3}\left(\frac{m_{\chi} H_0}{M_P^2}\right)\int_{q_0}^{\infty} \diff q\, q^2f_{\chi}(q)\,.
\label{eq:DMrelicgrav}
\end{equation}
Here, $q \equiv k/(a_{\rm end} H_{\rm{end}})$ is the rescaled dimensionless comoving momentum, with $q(k_{\rm{end}}) = 1$, $H_0 = 100 \, h \, \rm{km \, s^{-1} \, Mpc^{-1}}$ is the present Hubble parameter, $\rho_c = 1.05 \times 10^{-5} \, h^2 \, \rm{GeV \, cm^{-3}}$ is the present critical energy density, with $h \sim 0.67$~\cite{Planck:2018vyg}, and the present comoving scale (IR cutoff) is given by $q_0 \simeq 7 \times 10^{-30} (M_P^2/(H_{\rm{end}} T_{\rm{reh}}))^{1/3}$~\cite{Martin:2010kz,Liddle:2003as,Ellis:2015pla, Garcia:2022vwm}. We must ensure that the measured DM relic abundance satisfies the experimental value $\Omega_{\rm DM}h^2 = 0.1198$~\cite{Planck:2018vyg}, and it is obtained at some reheating temperature in the range $T_{\rm BBN}\leq T_{\rm reh}\leq T_{\rm{max}}$ with the BBN temperature of $T_{\rm BBN} \simeq 1~\text{MeV}$ and $T_{\rm{max}} = \left(90 H_{\rm{end}}^2 M_P^2/\pi^2 g_{\rm{reh}} \right)^{1/4}$, which is the theoretical maximum temperature, corresponding to the instantaneous conversion of the total inflaton energy density to radiation at the end of inflation.

\noindent {\bf Resonant Dark Matter Production.---} When the nonminimal coupling is large, with $\xi\gtrsim 10$, the mode equation~(\ref{eq:modes}) takes the form of a Mathieu equation, which features parametric instabilities. These instabilities give rise to exponential quasi-stochastic excitations of modes $q\sim 1-10$, as discussed further in the Supplementary Material. To account for such effects, one must rely on full numerical solutions to the mode equation. Our analysis incorporates parametric resonance effects up to $\xi\simeq70$. For larger values, namely $\xi>70$, such an extensive excitation corresponds to a dark matter energy density that exceeds one percent of the inflaton energy density. This would lead to the fragmentation of the inflaton condensate and backreaction on the scalar curvature $R$. To account for these effects, one must employ more advanced tools, such as lattice simulations, as explored in Ref.~\cite{Figueroa:2021iwm, Lebedev:2022vwf}. However, the lattice studies lie beyond the scope of this work and we do not consider them here.

\noindent {\bf Thermal Production of Dark Matter.---} Since the transition from a matter-dominated to radiation-dominated universe can be slow, one must also account for the thermal production of DM through the gravitational scattering of Standard Model (SM) particles from the thermal bath $\textrm{SM} + \textrm{SM} \rightarrow \chi + \chi$~\cite{Bernal:2018qlk, Mambrini:2021zpp, Clery:2021bwz, Clery:2022wib, Haque:2021mab}. In this case, dark matter is generated via freeze-in during the reheating phase. The dark matter density can be inferred by solving the system of coupled differential equations describing the inflaton decay and reheating,
\begin{align}
    \dot{\rho}_{\phi} + 3H \rho_{\phi} & = -\Gamma_{\phi}\rho_{\phi} \,, \\  \dot{\rho}_{r} + 4H \rho_{r} & = \Gamma_{\phi} \rho_{\phi}  \, , \\
    \dot{n}_{\chi} + 3 H n_{\chi} & = R_{\chi} \, ,
\end{align}
where $\rho_{\phi}$ and $\rho_{r}$ are the energy density of the inflaton and radiation, respectively, and $R_{\chi}$ is the temperature-dependent thermal production rate. By considering this rate to be subdominant compared to the inflaton decay rate, we have neglected the corresponding term accounting for DM production on the right-hand side of the second equation, only mildly affecting the radiation energy density. The radiation temperature can be estimated by solving the first two equations. The equation for the DM number density $n_{\chi}$ can be determined by solving the Boltzmann equation expressed in terms of the comoving number density $Y_{\chi} = n_\chi a^3$,
\begin{equation}
    \label{eq:boltzmann}
   \frac{\diff Y_{\chi}}{\diff a} \; = \; \frac{a^2 R_{\chi}(a)}{H(a)} \, .
\end{equation}
We find that the total thermal production rate induced by all gravitational scattering processes $\textrm{SM} + \textrm{SM} \rightarrow \chi + \chi$ is given by
\bea
&&
R_{\chi} \; = \; 
\frac{\pi ^3 (2560\xi (3 \xi -1) +3997)}{41472000} \frac{T^8}{M_P^4}\, ,
\label{R0a}
\eea
and the thermally-produced DM relic abundance is
\begin{equation}
\label{eq:thermalabund}
\Omega_{\rm{DM, \, \rm{thermal}}} \; \simeq \; 1.9 \times 10^{9} \, g_{\rm{reh}}^{-3/2} \, \left( \frac{M_P R_{\chi}(T_{\rm reh})}{T_{\rm{reh}}^5} \right) \left( \frac{m_{\chi}}{1 \, \rm{GeV}} \right) \, .
\end{equation}
We note that for couplings $\xi = 0$ and $\xi = 1/3$ the thermal rate is identical and reduces to the result found in~\cite{Clery:2021bwz}. When the nonminimal coupling is large, the rate becomes $R_{\chi}(T) \simeq \pi^3 \xi^2  T^8/ (5400 M_P^4)$. The detailed calculations are given in the Supplemental Material (SM).

\noindent {\bf Isocurvature Constraints.---} When a light scalar dark matter field (spectator field) is excited during inflation, it inevitably leads to large isocurvature perturbations~\cite{Chung:2004nh, Chung:2011xd}. The rapid increase in dark matter energy density is primarily driven by the quadratic fluctuations that substantially contribute to the variance $\langle \chi^2\rangle$~\cite{Chung:2004nh,Chung:2015pga,Ling:2021zlj,Redi:2022zkt}. Our analysis assumes no initial misalignment for the dark matter scalar field at the beginning of inflation, with $\langle \chi\rangle = 0$, (and $\langle \chi^2\rangle=0$)
~\cite{Starobinsky:1994bd,Garcia:2023awt}. Notably, when $\langle \chi\rangle = 0$, the dark matter inhomogeneities do not directly affect the curvature perturbation, and they can be treated as pure isocurvature fluctuations in the comoving gauge~\cite{Chung:2011xd,Chung:2004nh}. The second-order contribution to the isocurvature power spectrum is given by~\cite{Liddle:1999pr,Chung:2004nh,Ling:2021zlj}
\begin{equation}
\label{eq:PS}
\mathcal{P}_{\mathcal{S}}(k) \;=\; \frac{k^3}{2\pi^2\rho_{\chi}^2}\int \diff^3\bx \ \langle \delta\rho_{\chi}(\bx)\delta\rho_{\chi}(0) \rangle e^{-i \bk\cdot\bx}\,,
\end{equation}
where $\rho_{\chi}$ and $\delta\rho_{\chi}$ denote the DM energy density and its fluctuation, respectively. The current constraints on the isocurature power spectrum provided by ${\textit Planck}$ are $\beta_{\rm iso}\;\equiv\; \mathcal{P}_{\mathcal{S}}(k_*)/( \mathcal{P}_{\mathcal{R}}(k_*) + \mathcal{P}_{\mathcal{S}}(k_*) )\;<\;0.038$ at the 95\% C.L. for the pivot scale $k_*=0.05\,{\rm Mpc}^{-1}$, where $ \mathcal{P}_{\mathcal{R}}(k_*) = 2.1 \times 10^{-9}$ is the curvature power spectrum~\cite{Planck:2018jri}. This imposes an upper limit on the isocurvature power spectrum $\mathcal{P}_{\mathcal{S}}(k_*)\lesssim 8.3 \times 10^{-11}$. To understand the isocurvature suppression intuitively, one can compute the variance averaged over the superhorizon modes, with $\langle \chi^2 \rangle \sim H^4/m_{\rm{eff}}^2$, and regard such long-wavelength contribution as coherent oscillations of the DM field~\cite{Ema3, Tenkanen:2019aij}. Approximating that the field displacement at the end of inflation is given by $\sqrt{\langle \chi^2 \rangle}$, with the energy density given by $\rho_{\chi} \simeq \frac{1}{2} m_{\chi}^2 \langle \chi^2 \rangle$, the isocurvature power spectrum becomes proportional to $\mathcal{P}_{\mathcal{S}}(k_*) \sim H^2/\langle \chi^2 \rangle \sim m_{\rm{eff}}^2/H^2$; it becomes more suppressed as $\xi$, and in turn $m_{\rm{eff}}$, increases. However, to account for the full spectrum evolution including the transitions, one must evaluate numerically Eq.~(\ref{eq:PS}).

For purely gravitational dark matter production, the isocurvature constraints from \textit{Planck} require that $m_{\chi}^2 + 12 \xi H_*^2 \gtrsim H_*^2/4$, where $H_*$ is the Hubble scale at the horizon exit. For a minimal coupling ($\xi = 0$), 
this constraint becomes $m_{\chi} \gtrsim H_*/2$~\cite{Garcia:2023awt}. Assuming a very light bare dark matter mass, with $m_{\chi} \ll 2\sqrt{3 \xi} H_*$, we find that the isocurvature limits are always avoided for $\xi > 1/48$. This implies that even a very small value of a nonminimal coupling is sufficient to satisfy \textit{Planck} isocurvature limits, and the conformal coupling case $(\xi = 1/6)$ always satisfies the constraint. 

\noindent {\bf Lyman-$\alpha$ Forest Constraints.---} 
In contrast to the conventional WIMP scenario, light dark matter particles produced in an out-of-equilibrium state may possess a considerable pressure component, resulting in the suppression of overdensities on galactic scales and subsequently implying a cutoff in the matter power spectrum $\mathcal{P}(k)$ for scales $k$ larger than the free-streaming horizon wavenumber $k_{\rm H}(a)$. Lyman-$\alpha$ forest measurements constrain such cutoff scale to be around $k_\text{H}(a=1)> 15 \, h \,\text{Mpc}^{-1}$. This can be translated into a lower limit on the mass of a generic warm dark matter (WDM) candidate decoupling from the thermal bath  $m_\text{WDM}>m_\text{WDM}^{\text{Ly}\mbox{-}\alpha} \; \simeq \; (1.9-5.3)~\text{keV at 95 \% C.L.}$~\cite{Narayanan:2000tp,Viel:2005qj,Viel:2013fqw,Baur:2015jsy,Irsic:2017ixq,Palanque-Delabrouille:2019iyz,Garzilli:2019qki}

For DM produced in an out-of-equilibrium state, one can translate the Lyman-$\alpha$ bound into a constraint on the DM mass by matching the corresponding equation of state parameters. This process leads to the following bound~\cite{Ballesteros:2020adh, Garcia:2022vwm}
\begin{equation}
\label{eq:lyalphaconst}
 \,m_{\chi}^{\text{Ly}\mbox{-}\alpha} \;=\; m_{\rm WDM}^{\text{Ly}\mbox{-}\alpha} \left(\frac{T_{\star}}{T_{\rm WDM,0}}\right)\sqrt{\frac{\langle q^2\rangle}{\langle q^2\rangle_{\rm WDM}}}\,,
\end{equation}
where $T_{\rm WDM,0}$ is the WDM temperature saturating the dark matter abundance~\footnote{This quantity only depends on $m_{\rm WDM}^{\text{Ly}\mbox{-}\alpha}$, see Ref.~\cite{Ballesteros:2020adh} for details.}.  $T_{\star}=H_{\rm{end}}(a_{\rm end}/a_0)$ is the characteristic energy scale of the produced DM, and the normalized second moment of the dark matter phase space distribution function $f_{\chi}(q)$ defined by
\begin{equation}
    \langle q^2\rangle \, \equiv \, \dfrac{\int \diff q\, q^4 f_\chi(q) }{\int \diff q\, q^2 f_\chi(q) } \,.
    \label{eq:secondmoment}
\end{equation} 
For a WDM candidate, this quantity is approximately $\langle q^2\rangle_{\rm WDM} \simeq 12.93$. We argue that large values of nonminimal coupling $\xi$ allow the scalar dark matter to be very light, and the values of $\xi$ can be constrained by the structure formation limits, as detailed in our parameter space analysis below. 

\begin{figure}[t!]
\includegraphics[width = 1\columnwidth]{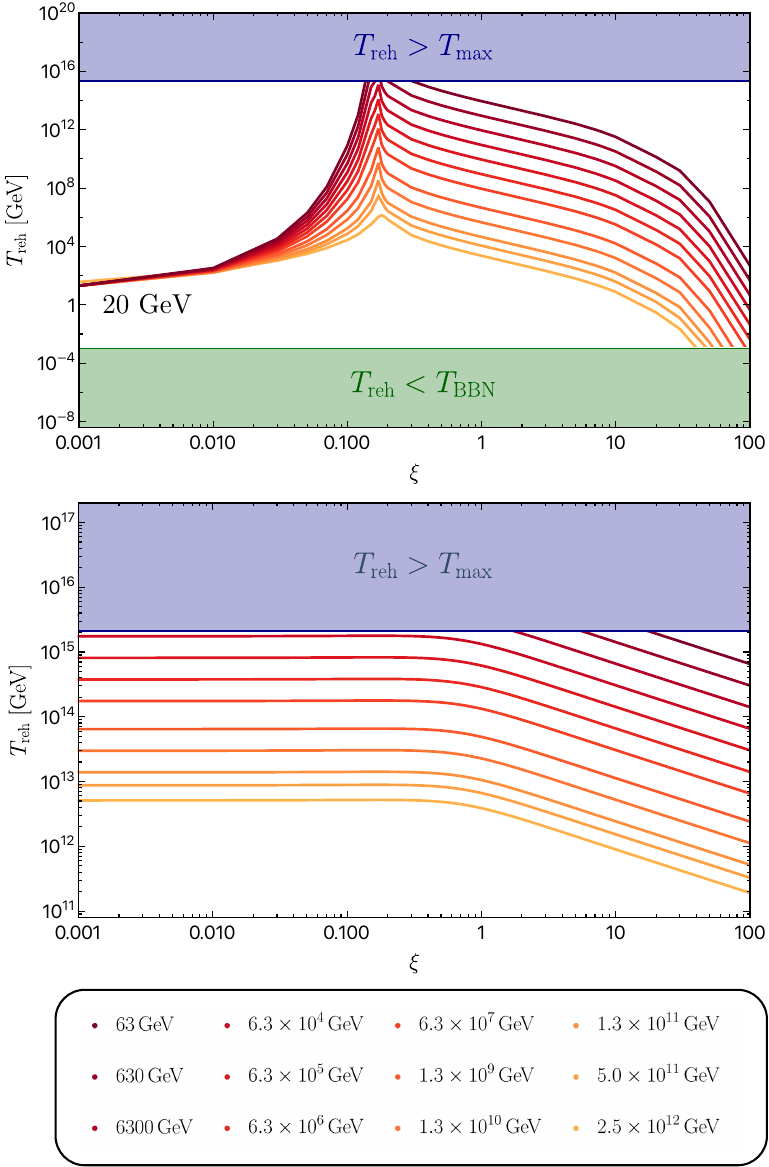}
\caption{Dependence of reheating temperature on $\xi$ for a range of masses. The top panel shows the constraints for the gravitational production, while the bottom panels illustrates the dependence for thermal production. The color coding represents the value of the bare dark matter mass $m_\chi$, as indicated in the legend. As evident from the plots, thermal production is always subdominant.}
\label{fig:tempplot}
\end{figure}

\begin{figure*}[!t]
\includegraphics[width = 0.8\textwidth]{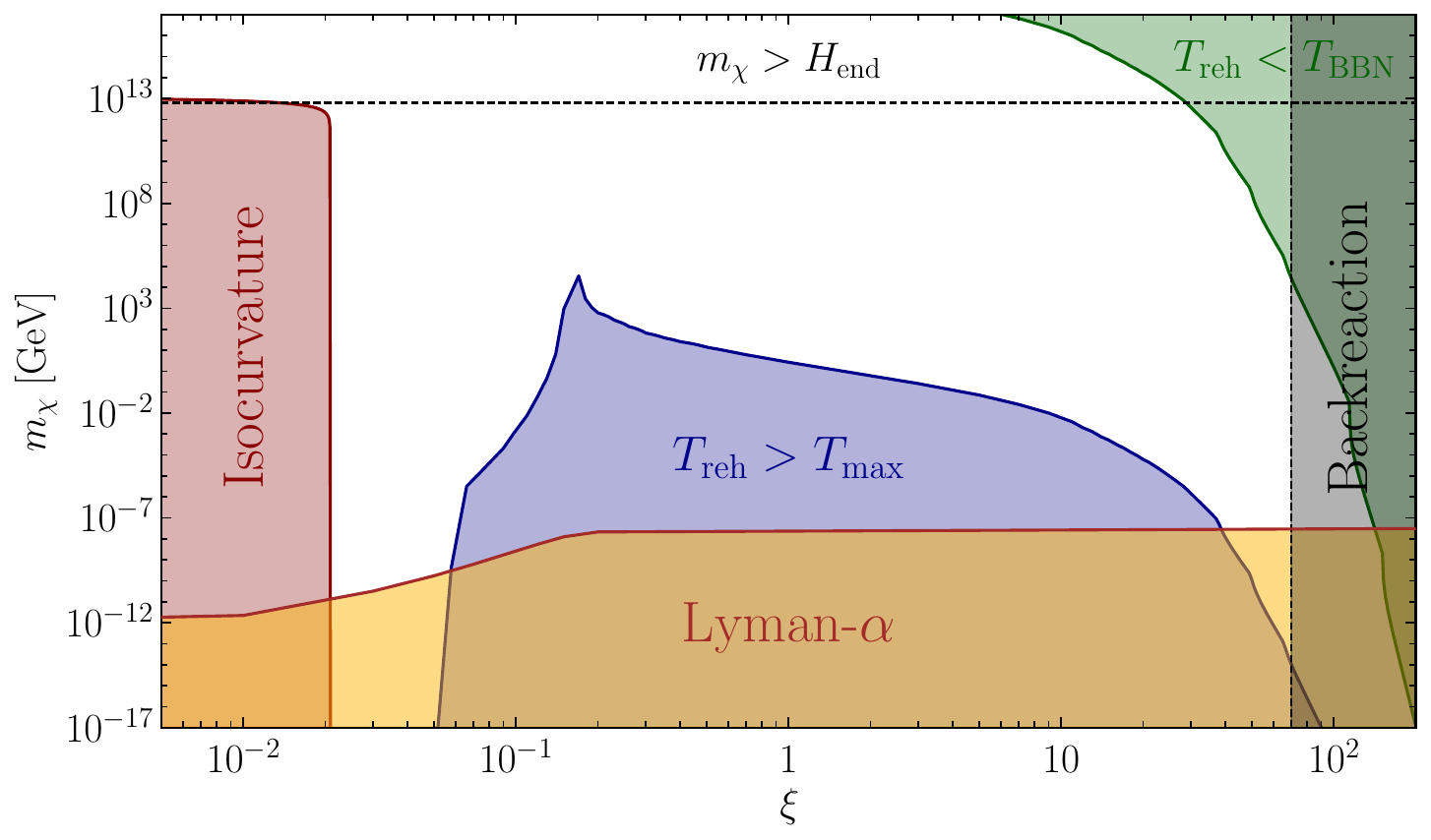}
\caption{Parameter space of the dark matter mass $m_{\chi}$ as a function of the nonminimal coupling $\xi$. The white region displays the space compatible with the observed dark matter abundance. The red region is ruled out by isocurvature constraints. The requirement for the reheating temperature to be less than the maximum temperature and greater than the BBN temperature, respectively, excludes the regions colored by blue and green. The yellow region is ruled out by Lyman-$\alpha$ constraints. The gray region indicates when the backreaction effects become important.}
\label{fig:parspace}
\end{figure*}

{\bf Results and Discussion.---} 
To impose the model constraints and demonstrate the effect of a nonminimal coupling $\xi$, we consider the T-model inflationary potential~\cite{Kallosh:2013maa}
\beq
\label{inf:tmodel}
V(\phi) \;=\; \lambda M_P^4 \left[ \sqrt{6}\tanh\left(\frac{\phi}{\sqrt{6}M_P}\right)\right]^2 \, ,
\eeq
where the potential can be normalized using the approximation $\lambda \simeq 3 \pi^2 A_{S*}/N_*^2$~\cite{Garcia:2020wiy}. For a nominal choice of $N_*=55$ $e$-folds, we obtain $\lambda \simeq 2 \times 10^{-11}$, the spectral tilt $n_s\simeq 0.963$, and the tensor-to-scalar ratio $r \simeq 0.004$, which is highly favored by current CMB measurements~\cite{Planck:2018jri, Ellis:2021kad}. The parameter $\lambda$ determines the inflaton mass at the potential minimum $V(0)$, with $m_{\phi} = \sqrt{2 \lambda} M_P \simeq 1.6 \times 10^{13} \, \rm{GeV}$. The Hubble parameter at the horizon exit is given by $H_* = 1.5 \times 10^{13} \, \rm{GeV}$ (with $\phi_* = 5.35 M_P$), and at the end of inflation, $H_{\rm{end}} = 6.3 \times 10^{12} \, \rm{GeV}$.

To determine how gravitational particle production depends on the nonminimal coupling $\xi$ and the reheating temperature $T_{\rm{reh}}$, we compute the DM relic density using Eq. (\ref{eq:DMrelicgrav}) and impose the observational value~$\Omega_{\rm DM}h^2 = 0.1198$~\cite{Planck:2018vyg}. Since the relic abundance depends on the reheating temperature, we apply the reheating temperature limits $T_{\rm BBN}\leq T_{\rm reh}\leq T_{\rm{max}}$, where the BBN temperature is $T_{\rm BBN} \simeq 1~\text{MeV}$ and $T_{\rm{max}} \simeq 2 \times 10^{15} \, \rm {GeV}$. We illustrate in the top panel of Fig.~\ref{fig:tempplot} the reheating temperature as a function of $\xi$ for a range of masses varying from $\mathcal{O}(10) \, \rm{GeV}$ to $\mathcal{O}(10^{12}) \, \rm{GeV}$. We find that when $m_{\chi}\ll H_{\rm end}$, as $\xi \rightarrow 0$ a universal limit is reached, with $T_{\rm{reh}} \simeq 20 \, \rm{GeV}$. As $\xi$ increases, the reheating temperature peaks in the range of $0.17 \lesssim \xi \lesssim 0.20$, and the conformal coupling $\xi = 1/6$ lies in this domain. Intuitively, this result can be better understood from Fig.~\ref{fig:genpsd}: when the nonminimal coupling is close to the conformal coupling value of $\xi = 1/6$, both the long-wavelength (IR) and the short-wavelength (UV) modes become suppressed in the comoving number density spectrum $\mathcal{N}_k$, and the total gravitationally-produced DM abundance is significantly reduced, which necessitates a very high reheating temperature to match the experimental value of $\Omega_{\chi} h^2$. However, as $\xi$ increases to larger values, the short-wavelength (UV) modes become the dominant contribution in the spectrum, and the required reheating temperatures now decrease exponentially. We find that for $\xi > 1$, the reheating temperature function can be fitted with the expression $T_{\rm{reh}}(m_{\chi}, \xi) = a  \exp(-b \, \xi) \left(1 \, {\rm{GeV}}/m_{\chi} \right)$, where $a = 1.1 \times 10^{14}$ and $b = 0.24$. 

Next, we use the thermal DM relic density expression~(\ref{eq:thermalabund}) and display $T_{\rm{reh}}(m_{\chi}, \xi)$ in the bottom panel of Fig.~\ref{fig:tempplot}. As can be observed, the thermally produced DM requires an extremely large reheating temperature close to $T_{\rm{max}}$. However, when comparing it with the purely gravitational particle production (top panel), we can see that the thermal production remains subdominant throughout the entire parameter space. Lastly, when evaluating cosmological parameters, it is important to adhere to the validity of the low-energy theory. The cutoff of this theory, which in this instance is the reheating temperature, can be approximated as $T_{\rm{reh}} \lesssim M_P/\xi$~\cite{Bezrukov:2010jz}. From our numerical analysis, we find that this constraint is always less stringent than the $T_{\rm{max}}$ constraint. 

To impose the isocurvature constraints, we numerically compute the isocurvature power spectrum~(\ref{eq:PS}) while imposing an experimental upper bound $\mathcal{P}_{\mathcal{S}}(k_*)\lesssim 8.3 \times 10^{-11}$~\cite{Planck:2018vyg}. We find that the fully numerical results are in excellent agreement with our given analytical approximations presented above. For minimal coupling, we find that $m_{\chi} \gtrsim 7.7 \times 10^{12} \, {\rm{GeV}} \simeq 1.2 \, H_{\rm{end}}$, which agrees with the well-known results that a scalar dark matter field with minimal coupling must be superheavy~\cite{Chung:2004nh, Chung:2011xd, Ling:2021zlj}; for a conformal coupling $\xi = 1/6$, we find that the isocurvature constraints are always satisfied, and in the massless limit we find that $\xi \gtrsim 0.02$. 

Finally, we estimate the Lyman-$\alpha$ bound~(\ref{eq:lyalphaconst}). We find that the dark matter bound increases as a function of $\xi$, with $m_{\chi} \gtrsim 2 \times 10^{-4} \, \rm{eV}$ when $\xi = 0$, and it plateaus around $\xi \gtrsim 0.2$, with the bound given by $m_{\chi} > 34 \, \rm{eV}$. The details of our derivation can be found in the Supplementary Material sections.

We show our combined parameter space in Fig.~\ref{fig:parspace}. In general, the nonminimal coupling $\xi$ is constrained to $0.1 < \xi < \mathcal{O}(100)$, with the lower bound resulting from the isocurvature constraint and the upper value from the BBN constraint. We note that we have not studied the superheavy region $m_{\chi} > H_{\rm{end}}$, which would also have constraints arising from $T_{\rm{BBN}}$ and $T_{\rm{max}}$, and we plan to investigate this in future work. We found that Lyman-$\alpha$ constraints give the lower bound $m_{\chi} > 2 \times 10^{-4} - 34 \, \rm{eV}$. The strongest constraint arises from $T_{\rm{max}}$ in the range of $0.1 \lesssim \xi \lesssim 1$.

In this Letter, we have explored a simple and compelling scenario involving a spectator DM scalar field that couples nonminimally to gravity. Our study demonstrates that the presence of nonminimal coupling opens up a broad parameter space, subject to constrains arising from maximum reheating temperature, BBN temperature, isocurvature, Lyman-$\alpha$ limits. We look forward to forthcoming experiments such as the Simons Observatory~\cite{SimonsObservatory:2018koc}, CMB-S4~\cite{Abazajian:2019eic}, and LiteBIRD~\cite{Hazumi:2019lys},
which could potentially detect B-modes in the CMB, provide more comprehensive scalar power spectrum analysis, and either strengthen isocurvature mode constraints or detect it. Since in this scenario, the DM field $\chi$ contains a blue-tilted isocurvature component, it predicts an enhancement in the power spectrum at small scales, making it amenable for being further constrained by improved structure formation and spectral distortion data~\cite{Newton:2020cog,Kogut:2011xw,Fu:2020wkq}. We plan to extend our study and examine the superheavy mass $m_{\chi} > H_{\rm{end}}$ regime and DM self-interaction effects in the upcoming work.   
\begin{acknowledgments}
We thank Mustafa Amin, Veronica Guidetti, Oleg Lebedev, Andrew Long, Konstantin Matchev, Michele Redi, and Jong-Hyun Yoon for helpful discussions. MG is supported by the DGAPA-PAPIIT grant IA103123 at UNAM, and the CONAHCYT ``Ciencia de Frontera'' grant CF-2023-I-17. MP acknowledges support by the Deutsche Forschungsgemeinschaft (DFG, German Research Foundation) under Germany's Excellence Strategy – EXC 2121 “Quantum Universe” – 390833306. The work of S.V. was supported in part by DOE grant DE-SC0022148. 
\end{acknowledgments}

\bibliography{references}

\clearpage

\onecolumngrid
\begin{center}
  \textbf{\large Supplementary Material for A New Window into Gravitationally Produced Scalar Dark Matter}\\[.2cm]
  \vspace{0.05in}
  {Marcos A. G. Garcia, Mathias Pierre, and Sarunas Verner}
\end{center}

\twocolumngrid
\setcounter{equation}{0}
\setcounter{figure}{0}
\setcounter{table}{0}
\setcounter{section}{0}
\setcounter{page}{1}
\makeatletter
\renewcommand{\theequation}{S\arabic{equation}}
\renewcommand{\thefigure}{S\arabic{figure}}
\renewcommand{\thetable}{S\arabic{table}}

\onecolumngrid

This Supplementary Material (SM) is organized as follows: In Sec.~\ref{sec:infandreheating}, we discuss the dynamics of slow-roll inflation and the reheating mechanism. Subsequently, in Sec.~\ref{sec:PSDs}, we demonstrate the computation of the phase space distributions (PSDs) for the T-model of inflation and calculate the purely thermal production of scalar dark matter. In Sec.~\ref{sec:isoc}, we present the analytical approximations of the isocurvature constraints. Lastly, in Sec.~\ref{sec:lya}, we discuss the full computation of the Lyman-$\alpha$ bound.

\section{Inflationary Dynamics and Reheating}
\label{sec:infandreheating}

\subsection{Slow-roll inflation}
In this section, we review the dynamics of the inflaton field, $\phi$, using the slow-roll approximation. We consider the following action:
\begin{equation}
    \mathcal{S} \; = \; \int \diff^4 x \sqrt{-g} \left[\frac{1}{2} (\partial_{\mu} \phi)^2 - V(\phi) \right] \, ,
\end{equation}
where the inflaton field is minimally coupled to gravity. Varying this action with respect to $\phi$ leads the Klein-Gordon equation of motion
\begin{equation}
    \ddot{\phi} + 3H \dot{\phi} + V' \; = \; 0 \, ,
\end{equation}
where the prime denotes a derivative with respect to the inflaton field, $\phi$. The Hubble rate is given by
\begin{equation}
    H^2 \; = \; \frac{1}{3M_P^2} \left[\frac{\dot{\phi}^2}{2} + V \right] \; \equiv \; \left( \frac{\dot{a}}{a} \right)^2 \, ,
\end{equation}
where $a$ is the scale factor. In our analysis, we employ the conventional single-field slow-roll inflationary parameters:
\begin{equation}
\label{eq:epseta}
\epsilon \equiv \frac{1}{2} M_{P}^{2}\left(\frac{V^{\prime}}{V}\right)^{2} \, , \qquad \eta \equiv M_{P}^{2}\left(\frac{V^{\prime \prime}}{V}\right) \, .
\end{equation}
By using the slow-roll approximation, we find that the number of $e$-folds can be expressed as
\begin{equation}
\label{eq:efolds}
N_{*} \simeq \frac{1}{M_{P}^{2}} \int_{\phi_{\mathrm{end}}}^{\phi_{*}} \frac{V(\phi)}{V^{\prime}(\phi)} \diff \phi \simeq \int_{\phi_{\mathrm{end}}}^{\phi_{*}} \frac{1}{\sqrt{2 \epsilon}} \frac{\diff \phi}{M_{P}} \, ,
\end{equation}
where the star denotes quantities at the horizon exit of $k_*  =  0.05 \, \rm{Mpc}^{-1}$, the pivot scale used in the {\it Planck} analysis. The end of inflation occurs when $\ddot{a} = 0$, which can be equivalently defined as $\dot{\phi}_{\rm{end}}^2 = V(\phi_{\rm{end}})$.

The primary CMB observables, specifically the scalar tilt, $n_s$, the tensor-to-scalar ratio, $r$, and the amplitude of the curvature power spectrum, $A_S$, can be expressed in terms of the slow-roll parameters 
as follows:
\begin{align}
    \label{eq:spectrtilt}
    n_{s} \; &\simeq \; 1-6 \epsilon_{*}+2 \eta_{*} \, , \\
    \label{eq:sclrtotens}
    r \; &\simeq \; 16 \epsilon_{*} \, , \\
    \label{eq:powerspectr}
    A_{S *} \; &\simeq \; \frac{V_{*}}{24 \pi^{2} \epsilon_{*} M_{P}^{4}} \, , 
\end{align}
where $V_* = V(\phi_*)$ and $A_{S *} \simeq 2.1 \times 10^{-9}$~\cite{Planck:2018vyg}. In our numerical analysis, we consider the T-model of inflation, given by Eq.~(\ref{inf:tmodel}). For this model, using the large $N_*$ limit, one can find the following cosmological observables~\cite{Ellis:2013nxa}
\begin{equation}
\label{cmbpredictions}
n_{s}  \; \simeq \; 1-\frac{2}{N_{*}}, \qquad r \; \simeq \; \frac{12}{N_{*}^{2}} \, .
\end{equation}
For a nominal choice $N_*=55$ $e$-folds, we find that the normalization constant is given by $\lambda \simeq 2 \times 10^{-11}$, the spectral tilt $n_s \simeq 0.964$, and the tensor-to-scalar ratio $r \simeq 0.004$. In this subsection, we have presented analytical approximations, and in the next subsection, we describe the full numerical procedure for the inflationary evolution and reheating.
\subsection{Reheating}
\begin{figure}[!t]
\includegraphics[width = 0.6\textwidth]{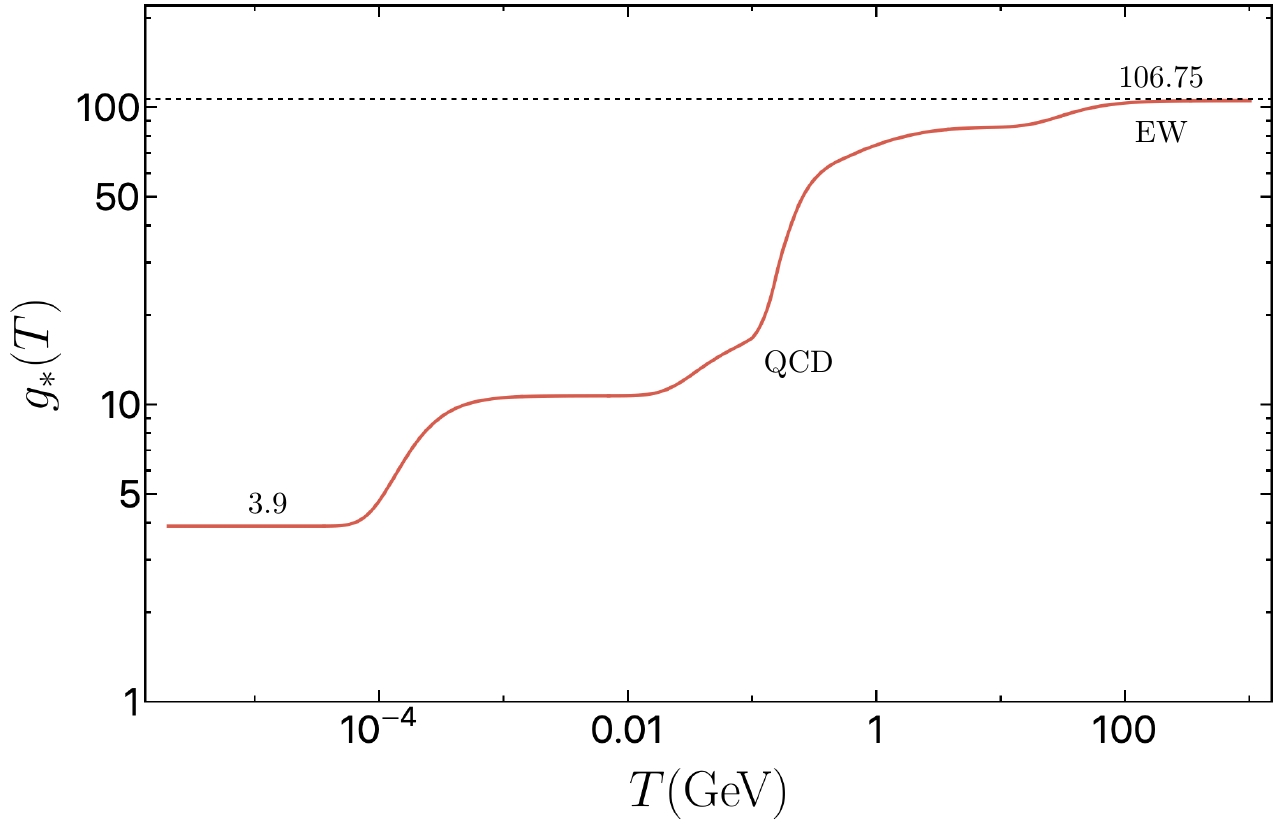}
\caption{The effective number of relativistic degrees of freedom in the Standard Model as a function of temperature.}
\label{fig:grehplot}
\end{figure}

In a more complete analysis, it is necessary to account for the duration of reheating, in order to accurately relate the number of $e$-folds of observable inflation with the {\em Planck} pivot scale. We thus compute the total number of $e$-folds, assuming no additional entropy production between the end of reheating and the time when the horizon scale $k_*$ reenters the horizon~\cite{Martin:2010kz, Liddle:2003as}:
\begin{equation}
\label{eq:nstarreh}
N_{*} \; = \; \ln \left[\frac{1}{\sqrt{3}}\left(\frac{\pi^{2}}{30}\right)^{1 / 4}\left(\frac{43}{11}\right)^{1 / 3} \frac{T_{0}}{H_{0}}\right]-\ln \left(\frac{k_{*}}{a_{0} H_{0}}\right) -\frac{1}{12} \ln g_{\mathrm{reh}}+\frac{1}{4} \ln \left(\frac{V_{*}^{2}}{M_{P}^{4} \rho_{\mathrm{end}}}\right) +\frac{1-3 w_{\mathrm{int}}}{12\left(1+w_{\mathrm{int}}\right)} \ln \left(\frac{\rho_{\mathrm{rad}}}{\rho_{\text {end }}}\right) 
\, .
\end{equation}
The present Hubble rate is given by $H_0 = 67.36\, \rm{km} \, \rm{s}^{-1} \, \text{Mpc}^{-1}$\cite{Planck:2018vyg} and the present photon temperature is $T_0 = 2.7255 \, \rm{K}$\cite{Fixsen:2009ug}. Here, $\rho_{\rm{end}}$ and $\rho_{\rm{rad}}$ are the energy density at the end of inflation and at the beginning of the radiation domination era when $w = p/\rho = 1/3$, respectively, $a_0 = 1$ is the present day scale factor, and $g_{\rm{reh}}$ represents the effective number of relativistic degrees of freedom in the Standard Model at time of reheating, typically taken to be $106.75$ for the reheating temperatures higher than $\mathcal{O}(100) \, \rm{GeV}$. We show the SM relativistic degrees of freedom $g_{\rm{*}}$ as a function of the temperature in Fig.~\ref{fig:grehplot}. We note, that in our full numerical analysis we considered the equation of motion parameter averaged over the $e$-folds during reheating, given by 
\begin{equation}
w_{\mathrm{int}} \equiv \frac{1}{N_{\mathrm{rad}}-N_{\mathrm{end}}} \int_{N_{\mathrm{end}}}^{N_{\mathrm{rad}}} w(n) \, \diff n \, ,
\end{equation}
where $N_{\rm{rad}}$ and $N_{\rm{end}}$ are the total number of $e$-folds at the end of inflation and at the beginning of full radiation domination era, respectively.

\begin{figure}[!t]
\includegraphics[width = 0.8\textwidth]{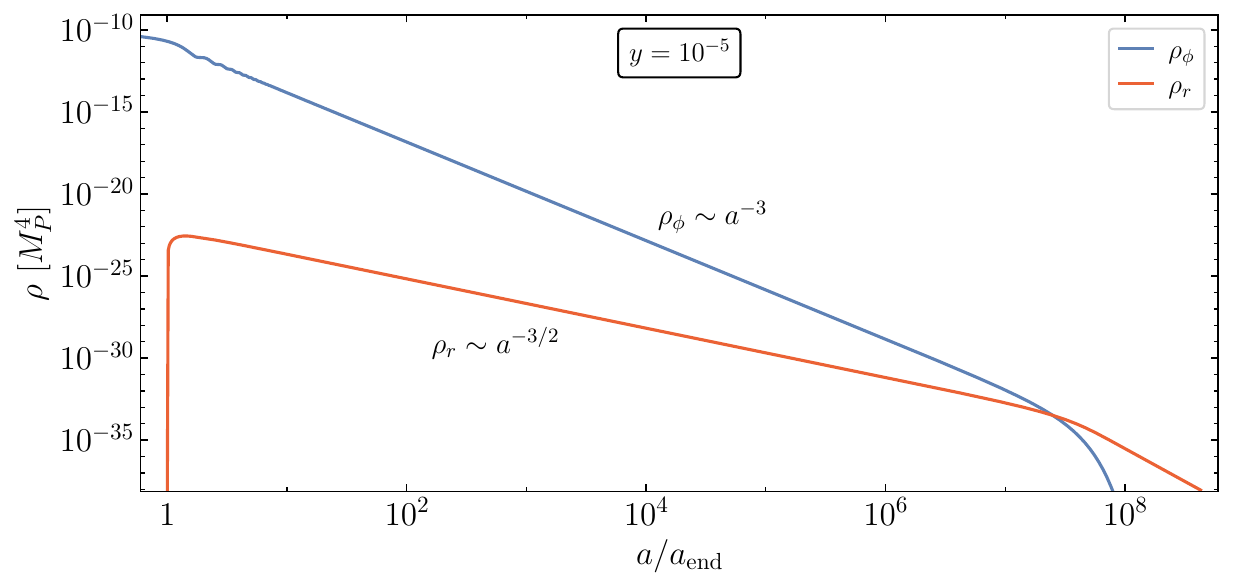}
\caption{Evolution of the energy densities during reheating of the inflaton field (blue) and radiation produced from the inflaton decay to fermions (red) for the Yukawa coupling of $y = 10^{-5}$. Reheating occurs when the two energy densities become equal to each other, $\rho_{\phi} = \rho_{r}$.}
\label{fig:endenplot}
\end{figure} 

The background dynamics are determined by the following system of coupled Friedmann-Boltzmann equations:
\begin{align}
\label{eq:dyn1}
\dot{\rho}_{\phi}+3 H \rho_{\phi}& =-\Gamma_{\phi} \rho_{\phi} \, , \\
\label{eq:dyn2}
\dot{\rho}_{r}+4 H \rho_{r} &=\Gamma_{\phi} \rho_{\phi} \, , \\
\label{eq:dyn3}
\rho_{\phi}+\rho_{r} &=3 M_{P}^{2} H^{2} \, ,\\
\frac{\diff}{\diff t}{(N w_{\rm int})} &= H w \,, 
\label{eq:dyn4}
\end{align}
where $\rho_{\phi}$ is the energy density of the inflaton. Here $\Gamma_{\phi}$ is the inflaton decay rate defined by
\begin{equation}
    \label{eq:decayrate}
    \Gamma_{\phi} \; = \; \frac{y^2}{8 \pi} m_{\phi} \, ,
\end{equation}
where $y$ is a Yukawa-like coupling. For T-models of inflation~(\ref{inf:tmodel}), we find that the inflaton mass is given by $m_{\phi} = \sqrt{2 \lambda} M_P \simeq 1.6 \times 10^{13} \, \rm{GeV}$, where $\lambda \simeq 2 \times 10^{-11}$. The universe undergoes reheating after the end of inflation, transitioning from a matter-dominated universe to a radiation-dominated universe. As the inflaton starts to decay, the thermal plasma dilutes until a maximum temperature of $T_{\rm{max}}$ is reached~\cite{Giudice:2000ex}. Subsequently, the temperature falls as $T \propto a^{-3/8}$, and the reheating temperature can be approximated as
\begin{equation}
    T_{\rm{reh}} \simeq 1.4 \times 10^{10} \, {\rm{GeV}} \cdot g_{\rm{reh}}^{-1/4} \left( \dfrac{y}{10^{-5}} \right)  \left(\frac{m_{\phi}}{1.6 \times 10^{13} \, {\rm{GeV}}} \right)^{1/2} \, .
\end{equation}

For the T-model of inflation~(\ref{inf:tmodel}), $N_{*}$ can be determined from the following simplified expression of the general expression~(\ref{eq:nstarreh})~\cite{Ellis:2021kad},
\beq
\label{eq:nstarfull}
N_{*} \;\simeq\; 58.36 - \frac{1}{2}\ln N_{*} + \frac{1}{6}\ln\left(\frac{ \Gamma_{\phi}}{m_{\phi}}\right) - \frac{1}{12}\ln g_{\rm reh} \, .
\eeq
Numerically, we find that inflation ends when $\phi_{\rm{end}} \simeq 0.84 \, M_P$ and $\rho_{\rm{end}} \simeq 7.1 \times 10^{62} \, \rm{GeV}^4$. We demonstrate the reheating process in Fig.~\ref{fig:endenplot} for the T-model of inflation and plot the evolution of energy densities during reheating with a nominal choice of the Yukawa coupling $y = 10^{-5}$.

\section{Computing the Scalar Particle Production and the PSDs}
\label{sec:PSDs}

\subsection{Perturbative gravitational production from the inflaton}
\label{sec:gravprod}

Since for gravitational production, DM annihilation, and thermalization can be safely neglected, the only relevant process for the determination of the PSD corresponds to the linear growth of the $\chi$ momentum modes. At the quantum field level, these modes are defined as in Eq.~(\ref{eq:modedef}), and satisfy the equation of motion (\ref{eq:modes}). To numerically solve for each mode during and after inflation, we use the attractor behavior of the background inflaton dynamics. We begin tracking the modes starting deep inside the horizon (in practice 5 $e$-folds prior to horizon crossing) with Bunch-Davies initial conditions, initial conditions for $\phi$ on the attractor, and unit scale factor at horizon crossing, $a=k/H=1$. This guarantees high precision in the numerical integration. The dark matter PSD can then be evaluated as the particle occupation number
\beq
\label{eq:occnum1}
f_{\chi}(k,t) =  \frac{1}{2\omega_k}\left| \omega_k X_k - i X'_k \right|^2\,.
\eeq
Here, $k$ corresponds to an arbitrarily defined comoving momentum. We find it convenient to refer this comoving momentum to the end of inflation, and introduce its dimensionless form,
\beq
q\;\equiv\; \frac{k}{a_{\rm end}H_{\rm end}} \;=\; \frac{K}{T_{\star}}\left(\frac{a}{a_0}\right)\,,\qquad T_{\star}\;\equiv\; H_{\rm end}\left(\frac{a_{\rm end}}{a_0}\right)\,,
\eeq
where $K$ denotes the physical momentum, and $T_{\star}$ is a characteristic energy scale. In this notation, the comoving DM number density is calculated as
\beq
n_{\chi}\left(\frac{a}{a_{\rm end}}\right)^3 \;=\; \frac{H_{\rm end}^3}{2\pi^2}\int \diff q\, q^2f_{\chi}(q)\,.
\eeq

\begin{figure}[!t]
\includegraphics[width = \textwidth]{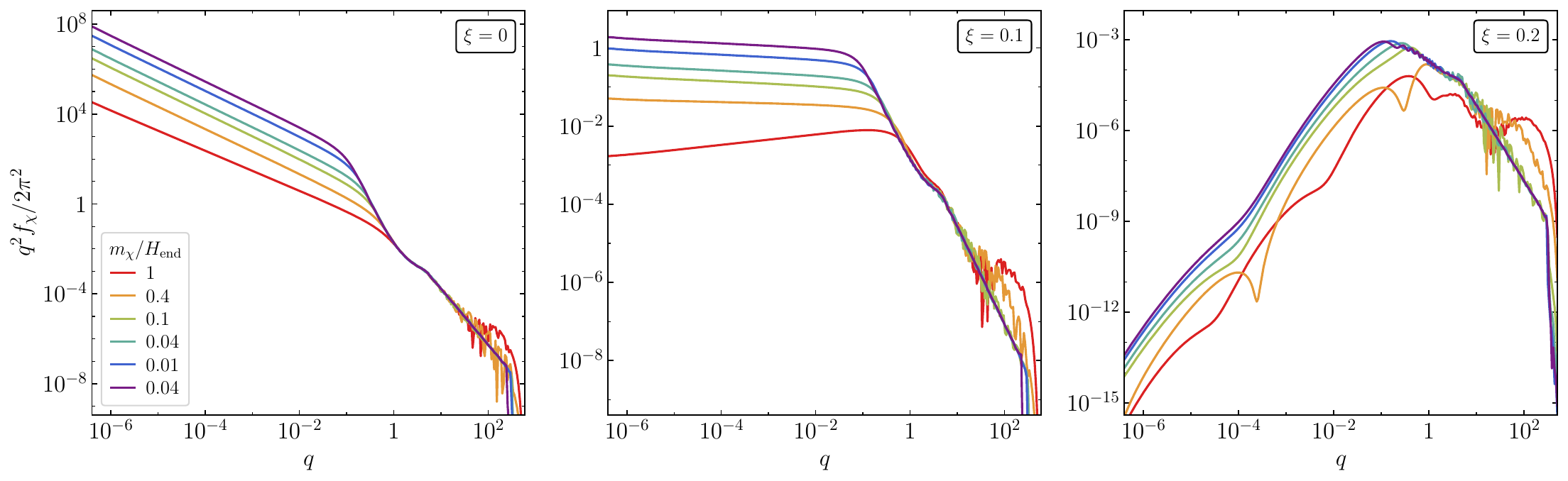}
\caption{Phase space distribution of a gravitationally produced DM particle $\chi$, for a selection of DM masses coded by color, and three different nonminimal couplings $\xi$. The PSDs are evaluated at $a/a_{\rm end}\simeq 130$.}
\label{fig:PSDs}
\end{figure}

Fig.~\ref{fig:PSDs} shows the numerically determined PSD for 3 different values of the nonminimal coupling $\xi$. For $\xi=0$, the distribution is red-tilted, and has a large amplitude in the IR for $m_{\chi}\ll H_{\rm end}$. This is the manifestation of the tachyonic de Sitter instability (see ``Framework'' in the main text). During inflation, modes with $k^2\lesssim -a^2m_{\rm eff}^2 \simeq 2a^2H^2$ have imaginary frequencies, which is translated into their rapid growth. In particular, modes that leave the horizon the earliest grow the most, leading to $f_{\chi}\propto q^{-3}$ in the IR~\cite{Herring:2019hbe, Ling:2021zlj, Kaneta:2022gug, Garcia:2022vwm}. As Fig.~\ref{fig:genpsd} also shows, a larger $\xi$ increases the effective mass of $\chi$, dampening the growth of the DM modes during inflation, leading to a blue tilt for $\xi>1/6$.

The IR endpoint of the PSD corresponds in principle to the mode that left the horizon at the beginning of inflation. This mode must satisfy $q_{\rm IR}\leq q_0$, where $q_0$ denotes the re-scaled present horizon scale. Following the standard computation that leads to (\ref{eq:nstarreh}), this scale may be written as~\cite{Martin:2010kz,Liddle:2003as,Ellis:2015pla}
\beq
q_0 \;\equiv\; \frac{a_0H_0}{a_{\rm end}H_{\rm end}} \;=\; \left(\frac{90}{\pi^2}\right)^{1/4}\left(\frac{11}{43}\right)^{1/3}\left(\frac{M_P}{H_{\rm end}}\right)^{1/2}\frac{H_0}{T_0} \frac{g_{\rm reh}^{1/12}}{R_{\rm rad}}\,,
\eeq
where~\cite{Ellis:2021kad}
\beq
R_{\rm rad} \;\equiv\; \frac{a_{\rm end}}{a_{\rm rad}}\left(\frac{\rho_{\rm end}}{\rho_{\rm rad}}\right)^{1/4} \;\simeq\; \left(\frac{\Gamma_{\phi}}{H_{\rm end}}\right)^{1/6}\,.
\eeq

\begin{figure}[!t]
\includegraphics[width = 0.8\textwidth]{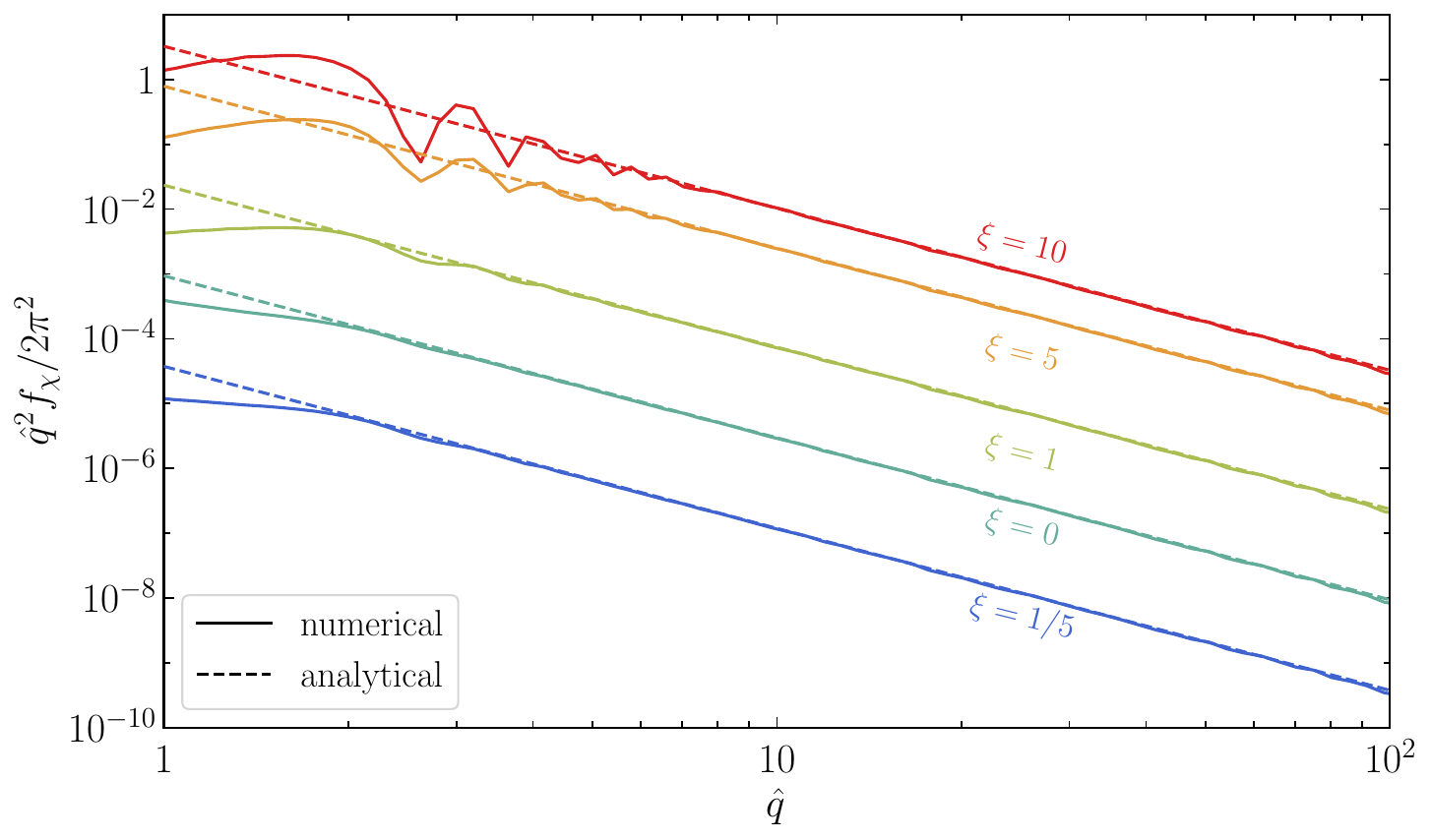}
\caption{Numerical and analytical evaluation of the phase space distribution as a function of the comoving momentum for selected values of $\xi$ and $m_\chi/H_\text{end}=10^{-3}$.}
\label{fig:PSDnumVSana}
\end{figure}

From Figs.~\ref{fig:genpsd} and \ref{fig:PSDs}, we note that for $q\gg 1$, corresponding to always sub-horizon modes, all PSDs for light DM scale similarly with $q$, although with $\xi$-dependent amplitudes. The modes are always particle-like, and their contribution to the PSD can be accurately estimated by integration of the Boltzmann equation for the process $\phi\phi\rightarrow \chi\chi$ via graviton exchange during reheating. The dark matter produced from the decay of the inflaton quanta of the oscillating coherent condensate follows the equation
\beq\label{eq:Boltzmannchi}
\frac{\partial f_{\chi}}{\partial t} - H|\boldsymbol{K}|\frac{\partial f_{\chi}}{\partial |\boldsymbol{K}|} \;=\; \frac{\pi|\mathcal{M}|^2}{2m_{\phi}^2}\delta(|\boldsymbol{K}|-m_{\phi})\,,
\eeq
in the case of a quadratic minimum of $V(\phi)$, $f_{\chi}\ll 1$, and $m_{\chi}\ll m_{\phi}$~\cite{Garcia:2022vwm}. The transition amplitude from the $\phi$ condensate to, whose definition can be found in Ref.~\cite{Garcia:2022vwm}, leads to
\beq
|\mathcal{M}|^2 \;=\; \dfrac{1}{8} \dfrac{\rho_\phi^2}{m_\phi^4} \lambda^2  \left( 1 - 6 \xi \right)^2  \, .
\eeq
Eq.~(\ref{eq:Boltzmannchi}) can be integrated approximately during reheating accounting for the conversion of $\rho_{\phi}$ into radiation. Denoting as $\theta$ the Heaviside step function, this solution is given by
\begin{align} \label{eq:fchipertq}
f_{\chi} \;&=\; \frac{\pi |\mathcal{M}(\hat{t})|^2}{2m_{\phi}^3H(\hat{t})}\theta(t-\hat{t})\theta (\hat{t}-t_{\rm end})\qquad \text{where} \qquad \frac{a(t)}{a(\hat{t})}\;=\; \frac{m_{\phi}}{|\boldsymbol{K}|}\\ \label{eq:fchipertq2}
&\simeq\; \frac{\sqrt{3}\pi \lambda^2 (1-6 \xi)^2 \rho_{\rm end}^{3/2}M_P}{16 m_{\phi}^7}\hat{q}^{-9/2}\times \begin{cases}
\theta\left( \frac{a(t)}{a_{\rm end}} -\hat{q} \right)\,, & t\ll t_{\rm reh}\\[7pt] 
e^{-1.56\left(\frac{a_{\rm end}}{a_{\rm reh}}\right)^2\hat{q}^2}\,, & t\gg t_{\rm reh}
\end{cases}
\end{align}
where $\hat{q}=(H_{\rm end}/m_{\phi})q$, for $q\gg 1$. The exponential tail is a phenomenological fit to (\ref{eq:fchipertq}) beyond the end of reheating. Due to the decay of the inflaton, $\rho_{\phi}(\hat{t}) \propto a^{-3}(\hat{t})e^{-\Gamma_{\phi} \hat{t}} \propto e^{-\kappa (|\boldsymbol{K}|/m_{\phi})^2}$ during radiation domination, where $\kappa$ must be numerically determined following the non-instantaneous transition from $w\approx0$ to $w\approx 1/3$.  Fig.~\ref{fig:PSDnumVSana} shows a comparison between the numerically evaluated PSD and the analytical approximation (\ref{eq:fchipertq2}) for modes excited between the end of inflation and the end of reheating. 

Having determined the PSD, the dark matter relic abundance is evaluated by integration of Eq.~(\ref{eq:DMrelicgrav}).

\subsection{Resonant gravitational production from the inflaton}
\label{sec:Backreaction}
\begin{figure}[!t]
\includegraphics[width = 0.48\textwidth]{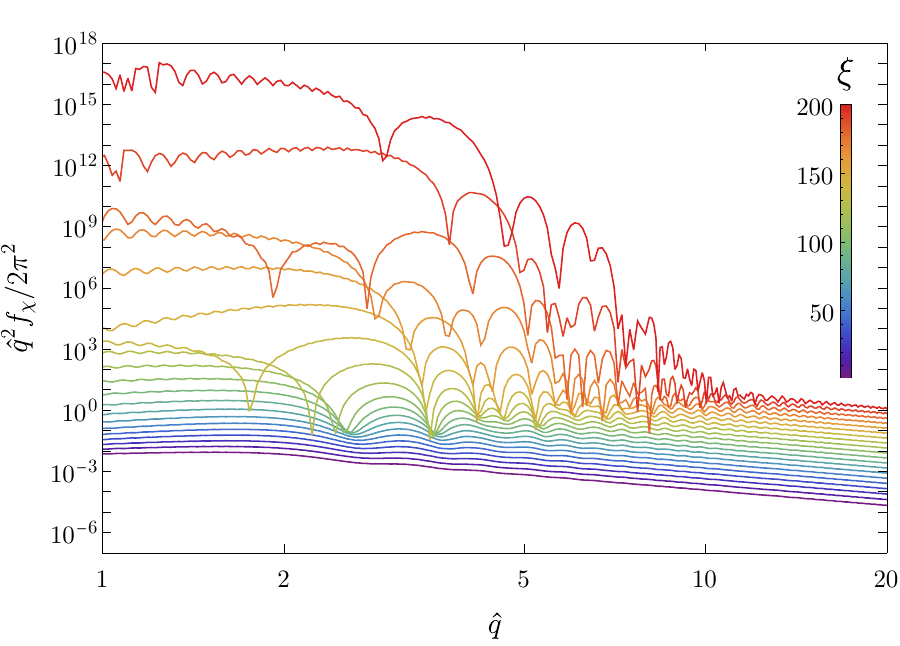} \hfill \includegraphics[width = 0.48\textwidth]{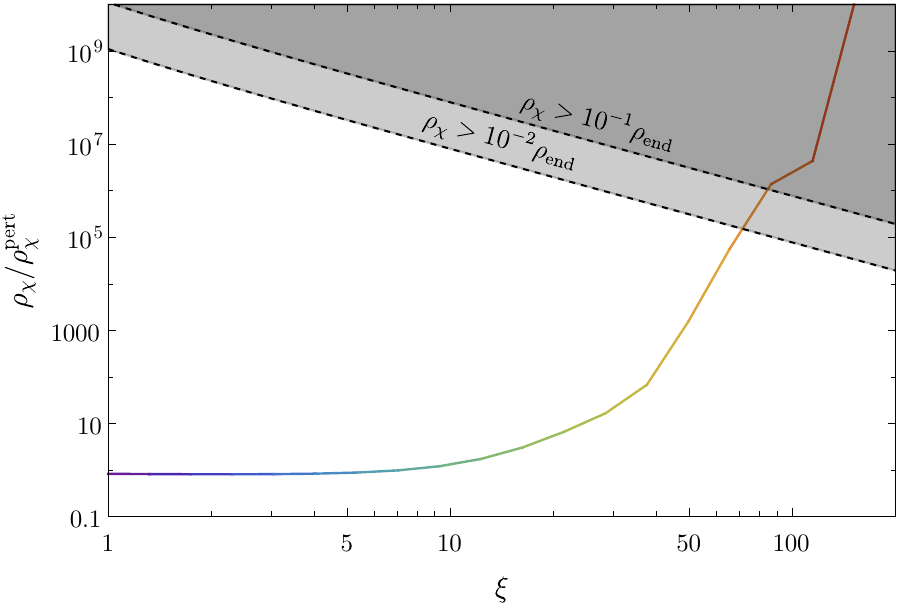}
\caption{Left: Numerical evaluation of the phase space distribution as a function of the comoving momentum for selected values of $1<\xi<200$ for $m_\chi/H_\text{end}=10^{-3}$. Right: Ratio of the total dark matter energy density obtained by numerically integrating the PSD with respect to the perturbative (Boltzmann) prediction obtained by integrating (\ref{eq:fchipertq2}).}
\label{fig:PSDnumresonances}
\end{figure}

For large values of $\xi \gg 1$, the scalar dark matter candidate is strongly coupled to the oscillating background, whose curvature can be expressed in terms of the inflaton field as
\begin{equation}
    R \, \simeq \, -\dfrac{1}{M_P^2}\big(4V(\phi)-\dot{\phi}^2\big) \,,
\end{equation}
assuming that the DM energy density is subdominant. After the end of inflation, the background inflaton field can be approximated by
\begin{equation}
    \phi(t) \, \simeq \, \phi_\text{end} \left( \dfrac{a}{a_\text{end}} \right)^{-3/2} \cos\big( m_\phi(t-t_\text{end}) \big) \,,
\end{equation}
which can be used to express the Ricci scalar as
\begin{equation}
    R \, \simeq \, - \dfrac{m_\phi^2 \phi_\text{end}^2}{2 M_P^2} \left( \dfrac{a_\text{end}}{a(t)} \right)^3 \Big( 1 + 3 \cos\big( 2 m_\phi(t-t_\text{end}) \Big) \Big) \,.
    \label{eq:Ricciapprox}
\end{equation}
Here we neglected the terms suppressed by powers of $H/m_\phi$ that decrease with time. The equation of motion for the rescaled field $\psi_k \equiv  a^{3/2} \chi_k$ takes the form of the Mathieu equation
\begin{equation}
    \dfrac{\diff^2 \psi_k}{\diff z^2}  + \big[ A_k+2 \kappa \cos(4 z) \big]  \psi_k \, = \, 0 \,,
\end{equation}
where $z \equiv m_\phi(t-t_\text{end})/2$, and 
\begin{equation}
    A_k \, \equiv \, \dfrac{4 k^2}{a^2 m_\phi^2} +  \dfrac{2  \xi\phi_\text{end}^2}{M_P^2}  \left( \dfrac{a_\text{end}}{a} \right)^3  , \qquad \kappa \equiv \dfrac{3 \xi \phi_\text{end}^2}{M_P^2} \left( \dfrac{a_\text{end}}{a} \right)^3\,,
    \label{eq:coeffsMathieu}
\end{equation}
in agreement with Ref.~\cite{Lebedev:2022vwf}. The Mathieu equation contains parametric instabilities. Solutions to this equation may undergo parametric resonances for large values of $\kappa \gg 1$, i.e., $\xi \gg 1$. Such resonances can be seen in the left panel of  Fig.~\ref{fig:PSDnumresonances}, where numerical evaluation of the PSD is depicted for chosen values of $\xi\in [1,200]$. For these values of $\xi$, the UV tail of the distribution consistently matches the perturbative prediction from Eq.~(\ref{eq:fchipertq2}), but substantial non-perturbative growth for $\hat q<10$ is apparent for $\xi>50$. 

Since the coefficients present in the Mathieu equation~(\ref{eq:coeffsMathieu}) have a strong scale factor dependence, the exponential amplification of mode functions typically shuts off rather swiftly after reheating begins, approximatley at $a \simeq 10 \, a_\text{end}$, equivalent to $\mathcal{O}(15)$ inflaton oscillations. However, for exceedingly large values of $\xi$, backreaction on the inflaton could induce fragmentation of the condensate. Additionally, the background value of the Ricci scalar would depart from the estimate of Eq.~(\ref{eq:Ricciapprox}) as the dark matter energy density contribution would need to be taken into account. In both cases, our approach, which treats the dark matter as a spectator field, would break down, and one would need to simulate the coupled system of inflaton and dark matter fields on the lattice, which lies beyond the scope of our work and is explored in Ref.~\cite{Lebedev:2022vwf}.

We evaluate the value of $\xi$ where such effects are expected to be significant by noticing that the bulk of the DM energy density is concentrated in modes with $\hat q\sim 1$, corresponding to modes substantially excited near the end of inflation. This value can therefore be estimated by requiring the dark matter energy density not to exceed $1\%$ of the inflaton energy density at the end of inflation. This corresponds to $\xi \lesssim 70$ for $\rho_\chi/\rho_\text{end}<0.01$ and $\xi \lesssim 85$ for $\rho_\chi/\rho_\text{end}<0.1$, as illustrated in the right panel of Fig.~\ref{fig:PSDnumresonances}, in agreement with Ref.~\cite{Lebedev:2022vwf}.

\subsection{Gravitational production from the Standard Model plasma}
\label{sec:thermalprod}
In this section we consider the production of nonminimally coupled scalar dark matter from scatterings between SM particles mediated by gravitational interactions. This contribution to the dark matter abundance can be estimated by solving the Boltzmann equation for the dark matter number density $n_\chi$
\begin{equation}
\frac{\diff n_\chi}{\diff t} + 3 H n_\chi = R_\chi\,,
\label{eq:Boltzmannrate}
\end{equation}
where $R_\chi$ is the thermal production rate that we derive in the following. \par \medskip

\noindent
\textbf{Interaction terms.} First, we expand the space-time metric around flat space $g_{\mu \nu} \simeq \eta_{\mu \nu} + 2h_{\mu \nu}/M_P$, where $h_{\mu \nu}$ represents the canonically-normalized (graviton) metric perturbation~\cite{Holstein:2006bh} and $\eta_{\mu \nu}$ the Minkowski metric. The relevant gravitational interactions are described by the following Lagrangian:
\begin{equation}
    \sqrt{-g} \mathcal{L}_{\rm{int}} \; = \; -\frac{h_{\mu \nu}}{M_P} \left(T^{\mu \nu}_{\rm{SM}} + T^{\mu \nu}_{\chi} \right) \, .
\end{equation}
Here $T^{\mu \nu}_{\rm{SM}}$ and $T^{\mu \nu}_{\chi}$ denote the stress-energy tensor of the Standard Model and the dark matter field, respectively. In general, the stress-energy tensor expression $T^{\mu \nu}_i$ depends on the spin of the field, $i = 0, \, 1/2, \, 1$. Assuming that all gravitational scattering processes take place at very high temperatures and neglecting masses of the SM particles, we can express the SM stress-energy tensor in the following form:
\begin{equation}
T^{\mu \nu}_{\text{SM}} \, \equiv \, \sum_{i=0,1/2,1}   T^{\mu \nu}_{{\text{SM}}, i}
\end{equation}
where each component can be written explicitly as
\bea
T^{\mu \nu}_{{\text{SM}}, 0} &=&
2(D_\mu H^\dagger)(D_\nu H) - g_{\mu \nu}
\left[D^\alpha H^\dagger D_\alpha H  \right] \,,
\label{Eq:tensors}
\\
T^{\mu \nu}_{{\text{SM}}, 1/2} &=&\sum_{\psi}
\frac{i}{4}
\left[
\bar \psi \gamma^\mu \overset{\leftrightarrow}{\partial^\nu} \psi
+\bar \psi \gamma^\nu \overset{\leftrightarrow}{\partial^\mu} \psi \right] -g^{\mu \nu}\left[\frac{i}{2}
\bar \psi \gamma^\alpha \overset{\leftrightarrow}{\partial_\alpha} \psi \right] \, , \\
\label{Eq:tensorf}
T_{{\text{SM}}, 1}^{\mu \nu} &=& \sum_{A_\mu} \frac{1}{2} \left[ F^\mu_\alpha F^{\nu \alpha} + F^\nu_\alpha F^{\mu \alpha} - \frac{1}{2} g^{\mu \nu} F^{\alpha \beta} F_{\alpha \beta} \right] \, ,
\label{Eq:tensorv}
\eea
where  $\psi$ represents a generic SM fermion, $A_\mu$ a SM gauge field with corresponding field strength tensor $F_{\mu \nu}$ and $H$ is the SM Higgs doublet with $D^\mu$ being the corresponding covariant derivative. We used the symbol $ \overset{\leftrightarrow}{\partial_\mu} \equiv \overset{\rightarrow}{\partial_\mu}-\overset{\leftarrow}{\partial_\mu}$. The sums are performed over all SM fields. Non-abelian indices are omitted for clarity. Terms giving rise to scattering processes involving more than two initial states were discarded.\\

With our conventions, we define the dark matter stress-energy tensor by $\sqrt{-g}T_{\mu \nu,\chi} = 2 \delta(\sqrt{-g} \mathcal{L}_{\chi})/\delta g^{\mu \nu}$~\cite{Birrell:1982ix}, which can be expressed as\footnote{This expression agrees with Ref.~\cite{Lebedev:2022vwf}.}
\begin{multline}
    T_{\mu \nu}^{\chi}  = \left(1 - 2 \xi \right) \left(\nabla_{\mu} \chi \right) \left(\nabla_{\nu} \chi \right) + \frac{1}{2} \left(4\xi - 1 \right) g_{\mu \nu}g^{\rho \sigma} \left(\nabla_{\rho} \chi \right) \left(\nabla_{\sigma} \chi \right) \\ - 2 \xi \left(\chi \nabla_{\mu} \partial_{\nu}\chi - g_{\mu \nu} \chi \Box{\chi} \right) + \xi \left[R_{\mu \nu } - \frac{1}{2} R g_{\mu \nu} \right] \chi^2 + \frac{1}{2}m_{\chi}^2 g_{\mu \nu} \chi^2 \, .
\end{multline}
with $\Box \equiv g_{\mu \nu} \nabla^\mu \nabla^\nu$ and $\nabla^\mu$ being the covariant derivative in curved space-time. Around the flat Minkowski limit, this expression reduces to
\begin{equation}
    T_{\mu \nu}^{\chi}  = \left(1 - 2 \xi \right) \partial_{\mu} \chi \partial_{\nu} \chi + \frac{1}{2}(1-4 \xi) \eta_{\mu \nu} \left(m_{\chi}^2 \chi^2 - \partial_{\alpha} \chi \partial^{\alpha} \chi \right) - 2 \xi  \chi \partial_{\mu} \partial_{\nu}\chi \, .
\end{equation}

\noindent
\textbf{Scattering amplitudes.}
The gravitational scattering amplitudes of the dark matter production process
${\rm{SM}}^i(p_1)+{\rm{SM}}^i(p_2) \rightarrow \chi(p_3) + \chi(p_4)$ can be parametrized by
\begin{equation}
\mathcal{M}_{i} \propto \dfrac{1}{M_P^2} M_{\mu \nu}^i \Pi^{\mu \nu \rho \sigma} M_{\rho \sigma}^{\chi} \;, 
\end{equation}
where $i = 0,1/2,1$ represents the spin of the initial Standard Model states involved in the scattering process.  $p_1, p_2$ and $p_3, p_4$ denote respectively the incoming and outgoing momenta. In this case, $\Pi^{\mu \nu \rho \sigma}$ denotes the propagator for the canonically-normalized graviton $h_{\mu \nu}$ carrying momentum $k = p_1 + p_2$~\cite{Giudice:1998ck}
\begin{equation}
 \Pi^{\mu \nu \rho \sigma}(k) = \frac{\eta^{\mu \rho} \eta^{\nu \sigma} + \eta^{\mu \sigma} \eta^{\nu \rho} - \eta^{\mu \nu} \eta^{\rho \sigma}}{2k^2} \, .
\end{equation} 
The $M_{\mu \nu}^{\chi}$ partial amplitude can be expressed as
\beq
M_{\mu \nu}^{\chi} =\frac{1}{2}\left[\left(1 - 2\xi \right)\left( p_{3\mu} p_{4\nu} + p_{3\nu} p_{4\mu} \right)- (1-4\xi)\eta_{\mu \nu} \left( p_3\cdot p_4 + m_{\chi}^2 \right) -2\xi \left(p_{3\mu}p_{3 \nu} + p_{4\mu}p_{4 \nu}\right)\right] \,.
\eeq

The SM partial amplitudes, $M_{\mu \nu}^i$, are given by
\bea 
M_{\mu \nu}^{0} &=& \frac{1}{2}\left[p_{1\mu} p_{2\nu} + p_{1\nu} p_{2\mu} - \eta_{\mu \nu}p_1\cdot p_2 \right] \,, \\ 
M_{\mu \nu}^{1/2} &=&  \frac{1}{4} {\bar v}(p_2) \left[ \gamma_\mu (p_1-p_2)_\nu + \gamma_\nu (p_1-p_2)_\mu \right] u(p_1) \, , \\
M_{\mu \nu}^{1} &=& \frac{1}{2} \bigg[ \epsilon_{2}^{*} \cdot \epsilon_{1}\left(p_{1 \mu} p_{2 \nu}+p_{1 \nu} p_{2 \mu}\right) - \epsilon_{2}^{*} \cdot p_{1}\left(p_{2 \mu} \epsilon_{1 \nu}+\epsilon_{1 \mu} p_{2 \nu}\right) - \epsilon_{1} \cdot p_{2}\left(p_{1 \nu} \epsilon_{2 \mu}^{*}+p_{1 \mu} \epsilon_{2 \nu}^{*}\right)
\nonumber\\
&+&p_{1} \cdot p_{2}\left(\epsilon_{1 \mu} \epsilon_{2 \nu}^{*}+\epsilon_{1 \nu} \epsilon_{2 \mu}^{*}\right) +g_{\mu \nu}\left(\epsilon_{2}^{*} \cdot p_{1} \epsilon_{1} \cdot p_{2}-p_{1} \cdot p_{2} \, \epsilon_{2}^{*} \cdot \epsilon_{1}\right) \bigg]  \, ,
\label{partamp}
\eea

The corresponding full amplitudes including symmetry factors accounting for initial and final states are
\bea
&&
|\overline{{\cal M}_0}|^2 = \frac{1}{4M_P^4} \frac{\left(\xi  s^2+t (s+t)\right)^2}{s^2}\, ,
\\
&&
|\overline{{\cal M}_{1/2}}|^2 = -\frac{1}{4M_P^4}\frac{t(s +t )\left(s+2 t\right)^2 }{s^2} \, ,
\\
&&
|\overline{{\cal M}_{1}}|^2 = \frac{1}{2M_P^4} \frac{t^2(s+t)^2}{ s^2} \, .
\eea

\noindent
\textbf{Production rate.}
Next, we calculate the production rate for scalar dark matter, $\chi$, in the presence of nonminimal coupling $\xi$. This rate $R_\chi$ can be calculated using the following expression~\cite{Anastasopoulos:2020gbu,Brax:2020gqg,Brax:2021gpe}
\begin{eqnarray}
     R_\chi = 2\, \sum_{i=0,1/2,1}
     \frac{N_i}{1024 \pi^6}\int f_i(E_1) f_i(E_2) E_1 \diff E_1 E_2 \diff E_2 \diff \cos \theta_{12}\int |\overline{{\cal M}_{i}}|^2 \diff \Omega_{13} \,.   
    \label{Eq:totalrate}
\end{eqnarray}
with $f_i$ is the Fermi-Dirac $(i=1/2)$ or Bose-Einstein ($i=0,1$) distribution. Here the factor of two accounts for two dark matter particles produced per scattering, $N_i$ denotes the number of each SM species of spin $i$: $N_0=4$ for 1 complex Higgs doublet, $N_1=12$ for 8 gluons and 4 electroweak bosons, and $N_{1/2}=45$ for 6 (anti)quarks with 3 colors, 3 (anti)charged leptons and 3 neutrinos. The infinitesimal solid angle is given by
\begin{equation}
    \diff \Omega_{13}=2 \pi \diff \cos \theta_{13} \, ,
\end{equation}
with $\theta_{13}$ and $\theta_{12}$ being the angle formed by momenta ${\bf{p}}_{1,3}$ and ${\bf{p}}_{1,2}$, respectively. In the massless limit, one can express the amplitude squared in terms of Mandelstam variables, $s$ and $t$, which can be related to the angles $\theta_{13}$ and $\theta_{12}$ by
\begin{align}
    t\,= & \, \dfrac{s}{2}\left( \cos \theta_{13}-1 \right) \, ,  \\  s\, = & \,2E_1E_2 \left(1-\cos \theta_{12} \right) \, .
\end{align}
Substitution and intrgration of (\ref{Eq:totalrate}) yields the following rate
\bea
&&
R_{\chi}(T) \; = \; 
\frac{\pi ^3 (2560\xi (3 \xi -1) +3997)}{41472000} \frac{T^8}{M_P^4}\, .
\label{R0a1}
\eea
\noindent
\textbf{Dark matter production.} From the rate of Eq.~(\ref{R0a1}), we can solve the Boltzmann equation~(\ref{eq:Boltzmannrate}) and deduce the contribution to the dark matter abundance produced from the thermal bath
\begin{equation}
\Omega_{\rm{DM, \, \rm{thermal}}} \; \simeq \; 1.9 \times 10^{9} \, g_{\rm{reh}}^{-3/2} \, \left( \frac{M_P R_{\chi}(T_{\rm reh})}{T_{\rm{reh}}^5} \right) \left( \frac{m_{\chi}}{1 \, \rm{GeV}} \right) \, .
\end{equation}
For the parameter space of interest this contribution is always subdominant compared to gravitational production.

\section{Isocurvature Constraints}
\label{sec:isoc}
In this section, we provide a detailed analysis of the numerical evaluation of the isocurvature power spectrum. Furthermore, we present a comprehensive analytical derivation 
based on the approach outlined in Ref.~\cite{Garcia:2023awt}.
\par \medskip

\subsection{Numerical evaluation of the isocurvature constraints.}
The isocurvature power spectrum can be evaluated by using the following expression~\cite{Liddle:1999pr,Chung:2004nh,Ling:2021zlj, Garcia:2023awt}:
\begin{align}
\label{eq:Psp2}
\mathcal{P}_{\mathcal{S}}(k) \;=\; \frac{1}{\rho_{\chi}^2} \frac{k^3}{2\pi^2}\int \diff^3\bx \ \langle \delta\rho_{\chi}(\bx)\delta\rho_{\chi}(0) \rangle e^{-i \bk\cdot\bx} \;  =\; \frac{k^3}{(2\pi)^5\rho_{\chi}^2a^8}\int \diff^3\bp \ P_X(p,|\bp-\bk|)\,,
\end{align}
where the integrand is given by
\begin{align}
\label{eq:integrandisocurvature}
P_X(p,q) \; \equiv \; |X'_p|^2|X'_{q}|^2 + a^4m_{\chi}^4 |X_p|^2|X_{q}|^2
& + a^2m_{\chi}^2\left[ (X_pX_p^{\prime *})(X_{q}X_{q}^{\prime *}) + {\rm h.c.} \right]\,.
\end{align}
For light effective dark matter masses, $m_{\chi}\ll H_{\rm end}$, the integral of $P_X(p,q)$ features IR ($p\rightarrow 0$) and colinear ($|\bp-\bk|\rightarrow 0$) divergences. They are regularized with a lower momentum (IR) cutoff taken as the mode that left the horizon at the beginning of inflation. In principle, the power spectrum must be evaluated after the dark matter decoupling time, which in this case corresponds to the end of reheating. Nevertheless, the  argument of (\ref{eq:Psp2}) rapidly converges during reheating, and a few ($5-6$) $e$-folds are sufficient to obtain a reasonable approximation for the value of $\mathcal{P}_{\mathcal{S}}$. The integral (\ref{eq:Psp2}) is convergent in the UV. Further details can be found in Ref.\,\cite{Garcia:2023awt}. 

\begin{figure}[!t]
\includegraphics[width = 0.8\textwidth]{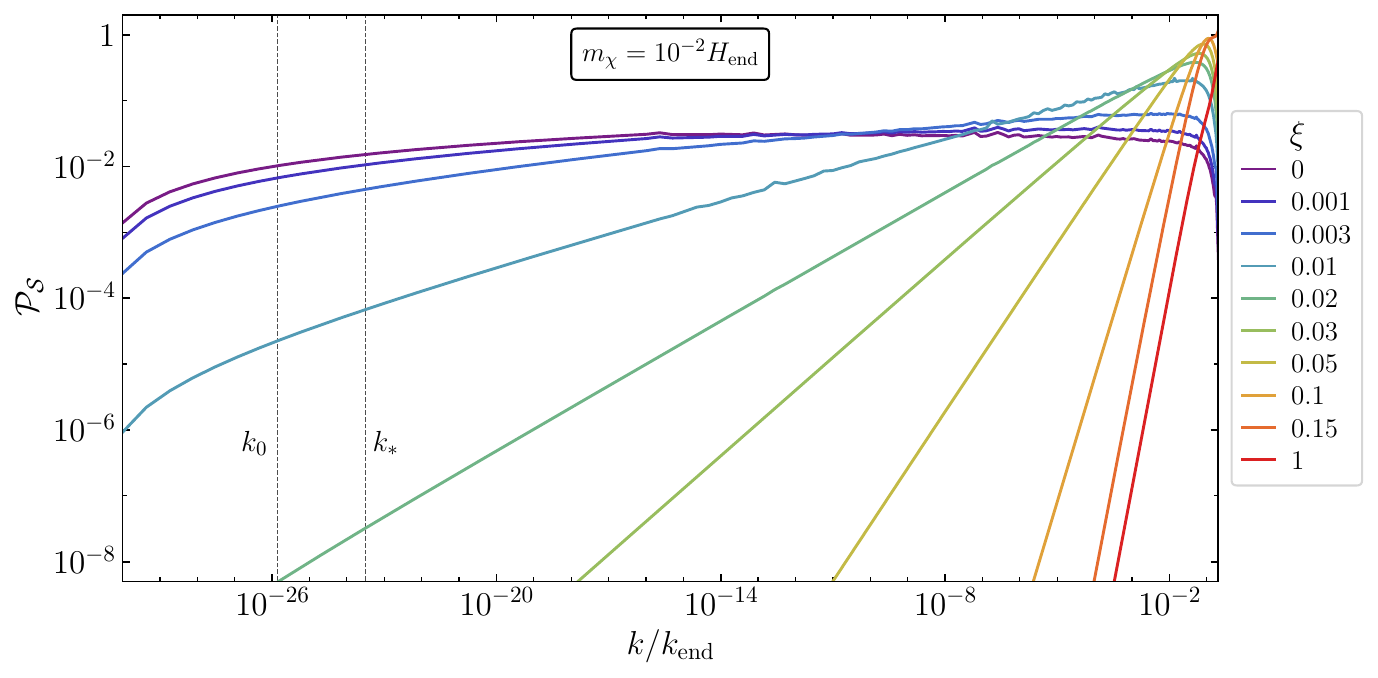}
\caption{DM isocurvature power spectrum for different nonminimal gravity-DM couplings with $m_{\chi}/H_{\rm end}=10^{-2}$, with each coupling represented by a different color. The vertical lines indicate the present horizon scale and the {\em Planck} pivot scale. Here we set $N_{\rm tot}=76.5$. The inflation model of choice corresponds to the T-model (\ref{inf:tmodel}).}
\label{fig:PS}
\end{figure}

Fig.~\ref{fig:PS} shows the numerically evaluated isocurvature power spectrum for different values of the nonminimal coupling $\xi$ for a light DM scalar with $m_{\chi}/H_{\rm end}=10^{-2}$. For this DM mass, the isocurvature constraint at the {\em Planck} pivot scale (right vertical dashed line) is averted for $\xi\gtrsim 0.03$. 

\subsection{Analytical estimate of the isocurvature constraints.} 
We model the dynamics of the universe during inflation as a de Sitter phase, which is described by a constant accelerated expansion of the universe, with $a(t) = e^{Ht}$ and a constant Hubble rate. We assume that this phase begins at an initial time $a(t_i) = a_i$ and concludes at $a(t_{\rm{end}}) = a_{\rm{end}}$. To determine the isocurvature power spectrum, we solve the mode equation for $\chi_k=  X_k/a$, expressed as a function of cosmic time and given by
\begin{equation}
    \ddot{\chi}_k+3 H \dot \chi_k +\left( \dfrac{k^2}{a^2} + m_\chi^2 + 12 \xi H^2 \right) \chi_k  \, = \, 0 \, ,
\end{equation}
where $m_\chi$ is the bare mass of the scalar field, which is assumed to be constant. Assuming Bunch-Davies initial condition and a pure de Sitter phase, the solution to the previous equation can be expressed as
\begin{equation}
    \chi_k(t) \, = \, \dfrac{\sqrt{\pi}}{2 a^{3/2}\sqrt{H}} e^{ i \frac{\pi}{2} (\nu+\frac{1}{2})} H_\nu^{(1)} \left(\dfrac{k}{aH}\right) \, ,
\end{equation}
where $H_\nu^{(1)} $ is the Hankel function of the first kind, and
\begin{equation}
    \nu \, \equiv \, \sqrt{\dfrac{9}{4} - 12\xi -\dfrac{m_\chi^2}{H^2}} \, .
\end{equation}
In the large wavelength limit $k \ll aH$, this solution reduces to
\begin{equation}
    \chi_k(k \ll aH) \, \simeq \, \dfrac{1}{2\sqrt{\pi} a^{3/2}\sqrt{H}} e^{ i \frac{\pi}{2} (\nu-\frac{1}{2})} \left( e^{-i \pi \nu}\Gamma(-\nu) \left( \dfrac{k}{2aH} \right)^{\nu} + \Gamma(\nu) \left( \dfrac{k}{2aH} \right)^{-\nu} \right)  \, .
\end{equation}
Neglecting the time derivative of the mode function and using the expression for the DM energy density valid at later times, $\rho_\chi \simeq m_\chi^2 \langle \chi^2 \rangle$, we can express Eq.~(\ref{eq:Psp2}) as:
\begin{align}
\mathcal{P}_{\mathcal{S}}(k) \; =\; \frac{k^2}{(2\pi)^4\langle \chi^2 \rangle^2}\int_0^{\infty} \diff  p\ p \int_{|k-p|}^{k+p}\diff q\ q \ |\chi_p|^2 |\chi_q|^2 \;  ,
\label{eq:isocurvatureanalytical}
\end{align}
where the dark matter field variance is
\begin{equation}
    \langle \chi^2 \rangle \, \equiv \, \int_{k_\text{IR}}^{k_\text{UV}} \, \mathcal{P}_\chi \, \diff \log k\, ,
\end{equation}
with
\begin{equation}
    \mathcal{P_\chi} \, \equiv \, \dfrac{k^3}{2 \pi^2}| \chi_k |^2 \, .
\end{equation}
Using the description above, we introduce a long-wavelength (IR) and short-wavelength (UV) cutoffs that regularize the power spectrum~\cite{Vilenkin:1983xp}, given by $k_\text{IR} = a_i H$ and $k_\text{UV}= a_{\rm{end}} H$. The total duration of de Sitter era is given by $N_\text{tot} \equiv \log(a_{\rm{end}}/a_i)$, where $\Delta N_* \sim 55 $ denotes the duration of inflation in $e$-folds between the CMB pivot scale $k_*$ crossing and the end of inflation. We discuss three distinct regimes in detail:
\par \medskip
\noindent
\textbf{Massless limit $m_\chi \ll H$.} 
If the DM mass can be neglected compared to the Hubble scale during the de Sitter phase, then $\nu = \sqrt{9/4 - 12 \xi}$, and the mode function can be expressed as
\begin{equation}
    | \chi_k(k \ll aH) | \, \simeq \, \frac{1}{\sqrt{2} \sqrt{a^3 H}} \left(\frac{k}{a H}\right)^{-\sqrt{9/4 - 12 \xi}} \, ,
    \label{eq:modereal}
\end{equation}
and the power spectrum is given by
\begin{equation}
    \mathcal{P}_{\chi} \; = \; \frac{H^{2}}{4 \pi^2} \left( \frac{k}{a H} \right)^{3 - \sqrt{9 - 48 \xi}}, \qquad \xi \leq \frac{3}{16} \, ,
\end{equation}
and it leads to a scale-invariant power spectrum $ \mathcal{P}_{\chi} = H^2/(4\pi^2)$ when $\xi = 0$ and $\mathcal{P}_{\chi} \; = \; H^{2}/(4 \pi^2) (k/(aH))^3$, when $\xi \geq 3/16$. If we combine expression~(\ref{eq:modereal}) with (\ref{eq:occnum1}), we find the the phase space distribution $f_{\chi}(k, t)$ for long-wavelength (IR) modes scales as $f_{\chi}(k,t)\sim |\chi_k|^2 \sim k^{-\sqrt{9 - 48 \xi}}$, which reduces to $f_{\chi} (k,t) \sim |\chi_k|^2 \sim k^{-3}$ for $\xi = 0$. For $\xi = 0$, at the end of inflation and the quasi-de Sitter phase, the variance can be approximated as
\begin{equation}
    \langle \chi^2 \rangle \, \simeq \,  \dfrac{H^2}{4 \pi^2} \log \left( \dfrac{a_{\rm{end}} }{a_i } \right) \, = \,\dfrac{H^2}{4 \pi^2} N_\text{tot}  \, .  
    \label{eq:variancesmallmass}
\end{equation}
 Evaluated at the CMB pivot scale, the isocurvature power spectrum~(\ref{eq:isocurvatureanalytical}) is
\begin{equation}
 \mathcal{P}^{m_\chi \ll H}_{\mathcal{S}}(k_*) \, \simeq \, \dfrac{N_\text{tot}-\Delta N_*}{N_\text{tot}^2} \, ,
 \label{eq:isocurvaturenegligiblemass}
\end{equation}
which scales as $1/N_\text{tot}$ for $\Delta N_* \ll N_\text{tot}$.\par \medskip

\noindent
\textbf{Limit $12 \xi + m_{\chi}^2/H^2 \lesssim 9/4$.} In this case $\nu \simeq 3/2 - \beta$, with $\beta \equiv m_{\chi}^2/(3H^2) + 4 \xi$ and the mode function takes the following form~\cite{Mukhanov:1990me, Liddle:1999pr}: 
\begin{equation}
    | \chi_k(k \ll aH) | \, \simeq \,  \dfrac{H}{\sqrt{2 a^3 H^3}}   \left(\frac{k}{a H}\right)^{-\nu } \,,
    \label{eq:moderealmedium}
\end{equation}
which leads to a slightly blue-tilted power spectrum
\begin{equation}
    \mathcal{P_\chi} \,= \dfrac{H^2}{4 \pi^2} \left( \dfrac{k}{aH} \right)^{ 2 \beta} \, .
\end{equation}
In this limit, the PSD $f_\chi(k,t)$ scales as $f_\chi(k,t)\sim |\chi_k|^2 \sim k^{-2 \nu} \sim k^{2\beta - 3}$. At the end of the quasi-de Sitter phase, the variance becomes 
\begin{equation}
    \langle \chi^2 \rangle \, \simeq \,  \left(\frac{1}{a H}\right)^{2 \beta } \frac{H^2}{8 \pi ^2 \beta }    \left(k_\text{UV}^{2 \beta }-k_\text{IR}^{2 \beta }\right)\, .     \label{eq:variancemediummass}
\end{equation}
When the quantity $\beta>10^{-2}$ and $k_\text{IR}= a_i H \simeq 10 ^{-10} k_*$ and $k_\text{UV}= a_{\rm{end}} H \simeq 10^{24} k_*$, the variance no longer depend on the long-wavelength (IR) modes. In this limit, one can approximate the power spectrum as
\begin{equation}
     \mathcal{P}^{m_\chi \lesssim H}_{\mathcal{S}}(k_*) \;\simeq \; \dfrac{ \mathcal{P}^{m_\chi \ll H}_{\mathcal{S}}(k_*)}{\mathscr{F}_0} \, \mathscr{F}\left( \beta \right) \,,
\label{eq:approximatedisocurvature}
\end{equation}
where
\begin{equation}
     \mathscr{F} (x) \, \equiv \, \dfrac{x^2 e^{-4 \Delta N_* x}}{16 \left(1-e^{-2 N_\text{tot} x}\right)^2}\,,
     \label{eq:curlyFfunction}
\end{equation}
and $\mathscr{F}_0=\lim_{x\rightarrow 0}\mathscr{F}(x)=1/(64N_\text{tot}^2)$. This expression implies that the isocurvature power spectrum decays exponentially. We note that the function $\mathscr{F} (x)$ is insensitive to $N_\text{tot}$ when $N_\text{tot} x \gg 1$. Therefore, the only non-negligible infrared sensitivity in Eq.~(\ref{eq:approximatedisocurvature}) arises from the scaling function $\mathcal{P}^{m_\chi \ll H}_{\mathcal{S}}(k_*) / \mathscr{F}_0\sim N_\text{tot}$ for a large $N_\text{tot}$ or equivalently $\sim \log(k/k_\text{IR})$ in terms of the infrared cutoff. \par \medskip 

\noindent
\textbf{Limit $12 \xi + m_{\chi}^2/H^2 \gtrsim 9/4$.} In this limit, $\nu \equiv -i \tilde{\nu} \in \mathbb{C}$ where $\tilde \nu \in \mathbb{R}^+$ and $\tilde \nu^2 \simeq 12 \xi+m_\chi^2 / H^2$. The mode function becomes
\begin{equation}
        \chi_k(k \ll aH) \, \simeq \,\dfrac{1}{2\sqrt{\pi} a^{3/2}\sqrt{H}} e^{  \frac{\pi}{2} (-\tilde \nu-\frac{1}{2}i)} \big[   e^{\pi \tilde \nu} \Gamma ( - i \tilde \nu ) e^{i \tilde \nu \log(x/2)}+ \Gamma ( i \tilde \nu ) e^{-i \tilde \nu \log(x/2)}         \big] \, ,
\end{equation}
where $x\equiv k/(aH)$ in the previous expression, and
\begin{equation}
     |   \chi_k(k \ll aH) |\, \simeq \,\dfrac{e^{-\pi \tilde \nu /2}}{2\sqrt{\pi} a^{3/2}\sqrt{H}} \, \mathcal{Z}(\tilde \nu)\, .
\end{equation}
Here $\mathcal{Z}(\tilde \nu)$ is given by
\begin{equation}
    \mathcal{Z}(\tilde \nu) \, \equiv \, \left(
    e^{\pi \tilde \nu} \Big[ 2 \cos \big(2 \tilde \nu \log(x/2) \big)  \Big( \mathcal{R}_{\tilde \nu}^2 -  \mathcal{I}_{\tilde \nu}^2 \Big) + 4 \mathcal{R}_{\tilde \nu} \mathcal{I}_{\tilde \nu} \sin \big(2 \tilde \nu \log(x/2) \big) \Big]
    + \left( 1 + e^{2 \pi \tilde \nu} \right) \left( \dfrac{\pi}{\tilde \nu \sinh (\pi \tilde \nu)} \right) \right)^{1/2} \, ,
\end{equation}
with $\mathcal{R}_{\tilde \nu}\equiv \text{Re}(\Gamma(i \tilde \nu))$ and $\mathcal{I}_{\tilde \nu}\equiv \text{Im}(\Gamma(i \tilde \nu))$. 
Since the first term in the squared brackets is an oscillatory term, with the oscillation frequency $2 \tilde \nu \log(k/(2aH))$, it can be ignored. Thus, in the limit $\tilde \nu \gg 1$, we obtain
\begin{equation}
    \mathcal{Z}(\tilde \nu \gg 1) \, \simeq \, \sqrt{\dfrac{2 \pi }{\tilde \nu}} e^{\pi \tilde \nu /2} \,.   
\end{equation}
This approximation works remarkably well even when $\tilde \nu \sim \mathcal{O}(1)$. The absolute value of the mode function can be approximated as
\begin{equation}
     |   \chi_k(k \ll aH) |\, \simeq \,\dfrac{1}{ a^{3/2}\sqrt{H}} \dfrac{1}{\sqrt{2 \tilde \nu}} \, \simeq \dfrac{1}{ \sqrt{2 m_\chi a^3}} \, ,
    \label{eq:modeimaginary}
\end{equation}
and the power spectrum becomes
\begin{equation}
    \mathcal{P_\chi} \, \simeq \, \dfrac{1}{4 \pi^2 a^3 m_\chi} k^3  \, \simeq  \dfrac{1}{4 \pi^2 m_\chi} H^3 e^{3(N_k-N)}\, \,.
\end{equation}
In the expression above, $N_k$ is the number of $e$-folds when the mode $k$ crosses the horizon $a(N_k)H=k$. The spectrum is convergent in the IR but exponentially decreases with the total number of $e$-folds after horizon crossing $N-N_k>0$,  and afterward it scales as non-relativistic matter. In this case, the field variance at the end of quasi-de Sitter phase becomes~\cite{Enqvist:1987au}
\begin{equation}
    \langle \chi^2 \rangle \, \simeq  \,  \dfrac{H^3}{12 \pi^2  m_{\rm{eff}}} \, ,
    \label{eq:variancelargemass}
\end{equation}
where we considered $m_{\rm eff} = \sqrt{m_{\chi}^2 - 2H^2(1-6\xi)}  \sim H$. In this limit, it becomes evident that the isocurvature power spectrum is exponentially suppressed:
\begin{equation}
 \mathcal{P}^{m_\chi \gtrsim H}_{\mathcal{S}}(k_*) \;\simeq \;   \frac{3 e^{-3 \Delta N_*}}{2} \, ,
 \label{eq:isocurvaturelargemass}
\end{equation}
and for a nominal choice of $\Delta N_*=55$, we find $\mathcal{P}^{m_\chi \gtrsim H}_{\mathcal{S}}(k_*) \simeq \mathcal{O}(10^{-70})$, a value substantially lower than both present and future constraints on isocurvature perturbations.\\

\noindent
\textbf{Estimate of the constraints.}
Using Eq.~(\ref{eq:isocurvaturelargemass}), one finds that the isocurvature constraint $\mathcal{P}_{\mathcal{S}}(k_*) < \beta_\text{iso} \mathcal{P}_{\mathcal{R}}(k_*) \simeq  10^{-11}$ is never violated when $m_\chi^2 > 9 H^2/4 - 12 \xi H^2$. In the massless limit with $\xi = 0$, as shown by Eq.~(\ref{eq:isocurvaturenegligiblemass}), the isocurvature power spectrum is extremely large, given by $ \mathcal{P}^{m_\chi \ll H}_{\mathcal{S}}(k_*) \, \sim \, \mathcal{O}(10^{-3})$ for $N_\text{tot} \, \sim \, \mathcal{O}(100)$, and the experimental constraint is always violated.

In the regime $12 \xi + m_{\chi}^2/H^2 \lesssim 9/4$, we can use the approximate expression~(\ref{eq:approximatedisocurvature}) to compute the analytical limits for the choice of parameter $\Delta N_* = 55$ and $N_{\rm{tot}} = 76.5$ $e$-folds. When $\xi = 0$, the limit becomes
\begin{equation}
m_{\chi}\, \gtrsim \, 0.57 \, H_* \, ,
\end{equation}
and in the massless limit, this bound becomes
\begin{equation}
\xi \, \gtrsim \, 0.027 \, ,
\end{equation}
which does not depend on the Hubble rate at the horizon exit, $H_*$. Importantly, this analytical approximation agrees remarkably well with the full numerical computation.

\section{Lyman-$\alpha$ Constraints}
\label{sec:lya}
In the standard $\Lambda$CDM model, dark matter is assumed to be a pressureless cold fluid. However, generic broad, non-thermal dark matter distributions, as obtained in the gravitational production scenario, could deviate from this $\Lambda$CDM picture. Such a deviation is constrained by the Lyman-$\alpha$ forest measurements of the matter power spectrum. In this section, we show the detailed derivation of such constraints for our nonminimally coupled dark matter model. \par \medskip

\noindent
\textbf{Non-cold dark matter.} For a generic dark matter candidate, the full space-time dependent distribution $\mathcal{F}_\chi(\bx,\bp,\eta)$ can be expanded in terms of a background quantity and a small fluctuation  $\Psi \ll 1$ as
\begin{equation}
\mathcal{F}_\chi(\bx,\bp,\eta) \, = \, f_\chi(\bp,\eta)(1+\Psi(\bx,\bp,\eta) ) \,,
\end{equation}
where $\bp$ is the momentum of DM particles located at the space coordinate  $\bx$. The  Fourier transform of the perturbed distribution $\Psi$ can be expanded in terms of Legendre polynomials $P_\ell$ as
\begin{equation}
\Psi(\bk,\boldsymbol{\hat{n}},q,\eta) \, = \, \sum_{\ell=0}^\infty (-i)^\ell (2\ell+1) \Psi_\ell (\bk,q,\eta) P_\ell	(\hat \bk \cdot \boldsymbol{\hat n}) \, ,
\end{equation}
where $k$ is the Fourier space wavenumber with $\bk = k \hat \bk$ and $q=\boldsymbol{\hat n} \cdot \bq$, where $\bq$ is the rescaled dimensionless dark matter momentum.
The Legendre coefficients $\Psi_\ell$ satisfy a system of Boltzmann equations involving metric perturbations as given in Ref.~\cite{Ballesteros:2020adh}. As argued in the same reference, terms with $\ell>1$ are typically suppressed for sufficiently cold relics, i.e.~for an EOS parameter $w_\chi \equiv \delta P_\chi / \delta \rho_\chi \ll 1$ defined as the ratio of the pressure to the  energy density perturbation for the DM species. In this case, this system of equations reduces to a coupled system of continuity and Euler equations for the dark matter relative energy-density perturbation $\delta_\chi \equiv \delta \rho_\chi/ \rho_\chi$ and the velocity perturbation. During matter domination, the equation for the dark matter overdensity becomes~\cite{Ballesteros:2020adh}
\begin{equation}
\ddot \delta_\chi + \mathcal{H} \dot \delta_\chi -\dfrac{3}{2} \mathcal{H}^2 \left( 1- \dfrac{k^2}{k_\text{FS}^2} \right) \delta_\chi \, = \, 0 \, ,
\end{equation}
where $\mathcal{H}\equiv a H$ is the conformal Hubble rate and $k_\text{FS}$ is the (time-dependent) free-streaming wave number given by
\begin{equation}
k_\text{FS} \, = \, \sqrt{\dfrac{9 \mathcal{H}^2}{10 w_\chi}} \,.
\end{equation}
This free-streaming wave number, uniquely determined by the EOS $w_\chi$, corresponds to a cutoff in the matter power spectrum as compared to the standard $\Lambda $CDM cosmology (with $w_\chi=0$) at the free-streaming horizon scale $k_\text{H}$, solely determined by $w_\chi$:
\begin{equation}
k_\text{H}(a) \, = \, \left( \int_0^a \dfrac{1}{k_\text{FS}(\tilde a)}  \dfrac{\diff \tilde a}{\tilde a} \right)^{-1}\,.
\end{equation}
The equation of state can be related to the normalized second moment of the DM background distribution
\begin{equation}
    \langle q^2\rangle \, \equiv \, \dfrac{\int \diff q\, q^4 f_\chi(q) }{\int \diff q\, q^2 f_\chi(q) } \,,
\end{equation} 
with
\begin{equation}
w_\chi \, \simeq \, \dfrac{T_\star^2}{3 m_\chi^2}\dfrac{\langle q^2 \rangle}{a^2} \,.
\label{eq:EOSchi}
\end{equation}

\noindent
\textbf{Lyman-$\alpha$ constraints.} Lyman-$\alpha$ forest measurements provide a bound on the cutoff scale to $k_\text{H}(a=1)> 15 \, h \,\text{Mpc}^{-1}$. This can be translated into a lower limit on the DM equation of state for a typical warm dark matter candidate
\begin{equation}
\label{eq:compareEOS}
w_\chi < w_\text{WDM}^{\text{Ly-}\alpha} \,.
\end{equation}
This corresponds to a constraint on the mass of a generic warm dark matter (WDM) candidate priorily in thermal equilibrium with the SM bath that decoupled while still relativistic $m_\text{WDM}>m_\text{WDM}^{\text{Ly}\mbox{-}\alpha} \; \simeq \; (1.9-5.3)~\text{keV at 95 \% C.L.}$~\cite{Narayanan:2000tp,Viel:2005qj,Viel:2013fqw,Baur:2015jsy,Irsic:2017ixq,Palanque-Delabrouille:2019iyz,Garzilli:2019qki}. From Eq.~(\ref{eq:EOSchi}) and Eq.~(\ref{eq:compareEOS}) one deduces the following bound for a dark matter candidate with a given distribution
\begin{equation}
 m_\chi\,>\,m_{\chi}^{\text{Ly}\mbox{-}\alpha} \;=\; m_{\rm WDM}^{\text{Ly}\mbox{-}\alpha} \left(\frac{T_{\star}}{T_{\rm WDM,0}}\right)\sqrt{\frac{\langle q^2\rangle}{\langle q^2\rangle_{\rm WDM}}}\,,
 \label{eq:lyalphaconst2}
\end{equation}
where $\langle q^2\rangle_{\rm WDM}\simeq 12.93$ $T_{\rm WDM,0}$ is the WDM temperature saturating the dark matter abundance. From Eq.~(\ref{eq:lyalphaconst2}), we get the following constraint for our setup
\begin{align}
m_\chi \, > \,  0.167 ~\text{keV}\, \left( \dfrac{m_{\rm WDM}^{\text{Ly}\mbox{-}\alpha}}{3~\text{keV}}\right)^{4/3} \left(\dfrac{m_\phi}{10^{13}~\text{GeV}} \right)^{1/3}  \left[ \left(\dfrac{T_\text{reh}}{M_P} \right)^{1/3}  \sqrt{\langle q^2 \rangle}\right]\, ,
\label{eq:Lymanalphaappendix}
\end{align}
where the terms inside the squared brackets depend on the bare dark matter mass and nonminimal coupling. The dependence of $T_\text{reh}$ on these parameters comes from the requirement for the dark matter to saturate the relic abundance.  \par \medskip

\noindent
\textbf{Constraints for} $\xi>1/4$.
In this case, as for the large direct DM-inflaton coupling regime of~\cite{Garcia:2022vwm}, the second moment $\langle q^2 \rangle$ is UV dominated and depends on the reheating temperature through
\begin{equation}
\langle q^2 \rangle \, \simeq \, 2.43 \, \sqrt{\dfrac{a_\text{reh}}{a_\text{end}}}\,.
\end{equation}
Interestingly, this scaling allows to cancel out the $T_\text{reh}$ dependence of Eq.~(\ref{eq:Lymanalphaappendix}) which asymptotes to a $\xi$-independent value for $\xi>1$ at
\begin{equation}
m_\chi \, > \, 32.4~\text{eV} \, \left( \dfrac{m_{\rm WDM}^{\text{Ly}\mbox{-}\alpha}}{3~\text{keV}}\right)^{4/3}  \left(\dfrac{m_\phi}{1.6 \times 10^{13}~\text{GeV}} \right)^{1/2} \, , \qquad  [\xi > 1/4] \,.
\label{eq:constraintmchiLyman}
\end{equation}	
Parametric resonances induce slight deviations from this relation based on the perturbative (Boltzmann) result. It is important to note that for $\xi>1$, the DM relic density scales as $\Omega_\chi h^2 \propto T_\text{reh}$. One can therefore rescale the reheating temperature by a factor $\alpha \equiv \rho_\chi/\rho_\chi^\text{pert}\geq 1$, such that the correct relic abundance is achieved. Importantly, this factor accounts for deviations from the perturbative (Boltzmann) prediction. Similarly, one can define a coefficient $\beta \equiv \langle q^2 \rangle/\langle q^2 \rangle^\text{pert}$. The constraints of Eq.~(\ref{eq:Lymanalphaappendix}) can thus be rescaled by an overall factor of $\alpha^{-1/3} \beta^{-1/2}$. Numerically, we find that such a factor does not significantly depart from unity for the range of interest $\xi\in[1,100]$. Therefore, for simplicity, we can approximate the Lyman-$\alpha$ bound by Eq.~(\ref{eq:constraintmchiLyman}) for large values of $\xi$ up to the backreaction regime $\xi \simeq 70$.

\noindent
\textbf{Constraints for} $\xi<1/4$. In this case, the PSD and corresponding second moment become sensitive to the IR part. In that case, Eq.~(\ref{eq:Lymanalphaappendix}) has to be evaluated numerically to account for the $T_\text{reh}$ and $\langle q^2 \rangle$ simultaneous dependence on $m_\chi$ and $\xi$. As the nonminimal coupling decreases, the Lyman-$\alpha$ bound approaches the limit for the pure gravitational production bound for a minimally coupled (e.g. $\xi =0$) dark matter candidate~\cite{Garcia:2022vwm}
\begin{equation}
m_\chi \, > 2 \times 10^{-4}~\text{eV}\, \qquad [\xi \ll 1] \, .
\end{equation}
The full numerical results are depicted in Fig.~\ref{fig:parspace}, that are found from a numerical evaluation of Eq.~(\ref{eq:Lymanalphaappendix}).

\end{document}